\documentclass[11pt,a4paper]{article}

\usepackage{apacite}  
\usepackage{epsfig}   
\usepackage{amssymb}  
\usepackage{amsmath}  
\usepackage{rotating} 

\newcommand{\Sec}[1]{Sec~\ref{sec:#1}}
\newcommand{\Fig}[1]{Fig~\ref{fig:#1}}

\newcommand{\eq}[1]{Eq~(\ref{#1})}
\newcommand{\Eq}[1]{\eq{eq:#1}} 
\newcommand{\Mean}[1]{\bar{#1}}
\newcommand{\Corr}[1]{\text{Corr}(#1)}
\newcommand{\Cov}[1]{\text{Cov}(#1)}
\newcommand{\tr}[1]{{#1}^{\mathrm{T}}}
\newcommand{\inv}[1]{{#1}^{\mathrm{-1}}}
\newcommand{\R}[1]{\mathbb{R}^{#1}}
\newcommand {\Dp}[2] {\frac{\partial #1}{\partial #2}}
\newcommand{\vek}[1]{\mathbf{#1}}
\newcommand{\sfcap}[1]{\caption{\textsf{#1}}}
\newcommand {\bmpsi}{\vek{\psi}}

\newcommand {\psm}{\Mean{\vek{\psi}}}
\newcommand {\psia}{\vek{\psi}^\mathrm{a}}
\newcommand {\psif}{\vek{\psi}^\mathrm{f}}
\newcommand {\xa}{\vek{x}_1}
\newcommand {\xb}{\vek{x}_2}

\newcommand{\bvec}[1]{\begin{bmatrix} \\ #1 \\ \\ \end{bmatrix}}

\newcommand{\incite}{\shortciteA}
\newcommand{\npcite}{\shortciteNP}

\renewcommand{\cite}{\shortcite}

\title{Real time assimilation of HF radar currents into a coastal ocean model}
\author{{{\O}yvind Breivik}\thanks{Presently at ECMWF, at the Nansen 
Environmental and Remote Sensing Centre, Bergen, Norway, at the time of
publication} \and {\O}yvind S{\ae}tra\thanks{The Norwegian 
Meteorological Institute, Oslo, Norway}}

\begin{document}

\date{First published March 2001}
\maketitle
\begin{abstract}
A real time assimilation and forecasting system for coastal currents is
presented. The purpose of the system is to deliver current analyses and
forecasts based on assimilation of high frequency radar surface current
measurements. The local Vessel Traffic Service monitoring the ship traffic to
two oil terminals on the coast of Norway received the analyses and forecasts
in real time.

A new assimilation method based on optimal interpolation is presented where
spatial covariances derived from an ocean model are used instead of simplified
mathematical formulations. An array of high frequency radar antennae provide
the current measurements. A suite of nested ocean models comprise the model
system. The observing system is found to yield good analyses and short range
forecasts that are significantly improved compared to a model twin without
assimilation. The system is fast, analysis and six hour forecasts are ready at
the Vessel Traffic Service 45 minutes after acquisition of radar measurements.
\end{abstract}

\noindent
Subject keywords: data assimilation, ocean modelling, current sensors, ocean
currents. \\
Regional terms: Europe, Norway, Bergen, Fedje. Bounding coordinates:
($4^\circ$~E, $60^\circ$~N), ($5^\circ$~E, $61^\circ$~N).\\
\emph{Journal of Marine Systems}, \textbf{28} (3--4), pp 161--182, 
doi:10.1016/S0924-7963(01)00002-1.

\section{Introduction}
Operational forecasting of the oceans is still in its infancy. Unlike the
atmospheric weather, ocean currents do not pose an immediate threat to everyday
life. Our daily goings-on may indeed never be affected by the strength of the
ocean currents in nearby waters. This explains in part why forecasting the
state of the atmosphere was a science and a craft as early as in the 1920s
while at the turn of the millenium the same can still not be said about the
oceans. (See the account of the first, failed attempt at numerically
forecasting the atmosphere in \npcite{ric22}.)

Another reason for this disparity may be the general impression that the ocean
is not as capricious as the atmosphere. While this may be true on some time
scales and in some regions, it is certainly not true for the swift coastal
surface currents that feed on the contrast in temperature and salinity between
neighbouring water masses. When such baroclinic currents are further
enforced by the wind field and join up with the tidal motion, the result can be
surface currents up to 2 m/s (4 knots) that loosely follow the coastline. These
currents vary in width and strength and can quickly break into whorls and
eddies of various sizes (typically less than 80 km, see \npcite{joh89} and
\npcite{ike89}). Such a chaotic dynamical system is unpredictable over longer
periods, hence data assimilation of observations is vital when trying to
forecast coastal currents.

The complexity of coastal current fields tells us that the pertinent horizontal
length scale of ocean dynamics, the Rossby radius of deformation (see
\npcite{gil82}), which in turn dictates the required horizontal resolution of
the numerical models, is vastly different from that of the atmosphere. In the
atmosphere, this scale determines the horizontal extent of low and
high pressure systems (the good and the bad weather), which is in the range of
several hundred kilometers. The equivalent scale in the ocean is only a few
tens of kilometers. Hence, resolving eddies in ocean models is only done at an
enormous computational cost which seriously limits the horizontal extent of the
model domain.

When trying to forecast coastal currents with the aid of data assimilation, the
Rossby length scale becomes an acute problem. In weather forecasting,
observations of the density field are made with a dense network of radio
soundings that resolves the vertical structure of the atmospheric fronts. The
density fronts associated with a mesoscale ocean cyclone may be less than a
kilometer in horizontal extent.  Hence, to achieve the same forecast skill for
mesoscale activity in the ocean as in the atmosphere, the vertical and
horizontal density structure of the ocean eddies and their associated fronts
must be resolved using an extremely dense network of vertical density
profilers.  Although technically possible (using salinity-temperature-depth
meters, CTD, or expendable bathythermographs, XBT), it is neither financially
nor operationally feasible to build such a dense coastal real time observation
network.

We see that there are numerical, instrumental, and socioeconomic reasons why
operational forecasting of the coastal ocean is lagging behind its atmospheric
counterpart. However, the appearance of huge oil tankers has caused currents to
become a factor to reckon. A tanker moves almost unaffected through bad weather
and rough seas, but coastal currents can seriously deflect its bearing. While
aiming toward narrow sounds, or maneuvering through channels and straits, it is
imperative to know the strength and direction of the currents. Fortunately,
there is now a growing awareness of the threat that coastal currents pose
to ship traffic and the ensuing pollution and potential loss of lives that
such disasters may cause.

There exists however an inexpensive alternative to in situ sampling of
oceanographic data. Shore based high frequency (HF) radars provide high
resolution surface current coverage in near real time on an extensive
observation grid at a very low cost. Although surface currents do not provide a
complete picture of a coastal current, it does provide valuable information on
its extent, direction, and magnitude. Using HF radars to map
coastal currents is by now a well tested technique (see \npcite{bar77} and
\npcite{bar78} for early accounts of the methodology) which has matured over
the years into reliable instruments.

One of the aims of the project EuroROSE (European Radar Ocean Sensing) is to
combine real time radar observations of surface currents with a suite of ocean
forecast models using an assimilation scheme to deliver real time analyses and
forecasts of coastal currents to the Vessel Traffic Services (VTS) in dangerous
regions with extensive ship traffic. The net site
\texttt{http://ifmaxp1.ifm.uni-hamburg.de/EuroROSE} provides more information
on the project as a whole.

Assimilating HF radar surface currents into models of the coastal ocean is a
relatively new approach to improving coastal current forecasts. The authors are
only aware of two other similar efforts. \incite{oke02} describes a data
assimilation system which utilizes a CODAR HF radar (see \npcite{bar77} for a
description of an early version of the radar) and the Princeton Ocean
Model to produce forecasts of the wind-driven, mesoscale shelf circulation off
the Oregon coast. The results are promising, including the vertical impact of
the surface currents. However, this system is only used for hindcast studies.
The Rutgers University has an ongoing assimilation effort associated with their
underwater laboratory LEO-15 \cite{gle00} off the coast of New Jersey.  During
five week campaigns in the summer months, they perform real time assimilation
of CODAR radar data into their Regional Ocean Model System (ROMS, see
\npcite{son94} for a description of an early version of the model) and produce
forecasts of coastal currents. More details of the system can be found at
\texttt{http://marine.rutgers.edu/mrs/}.

\begin{figure}[h]
   \begin{center}
   \centerline{
   \begin{tabular}{ll}
      \epsfig{file=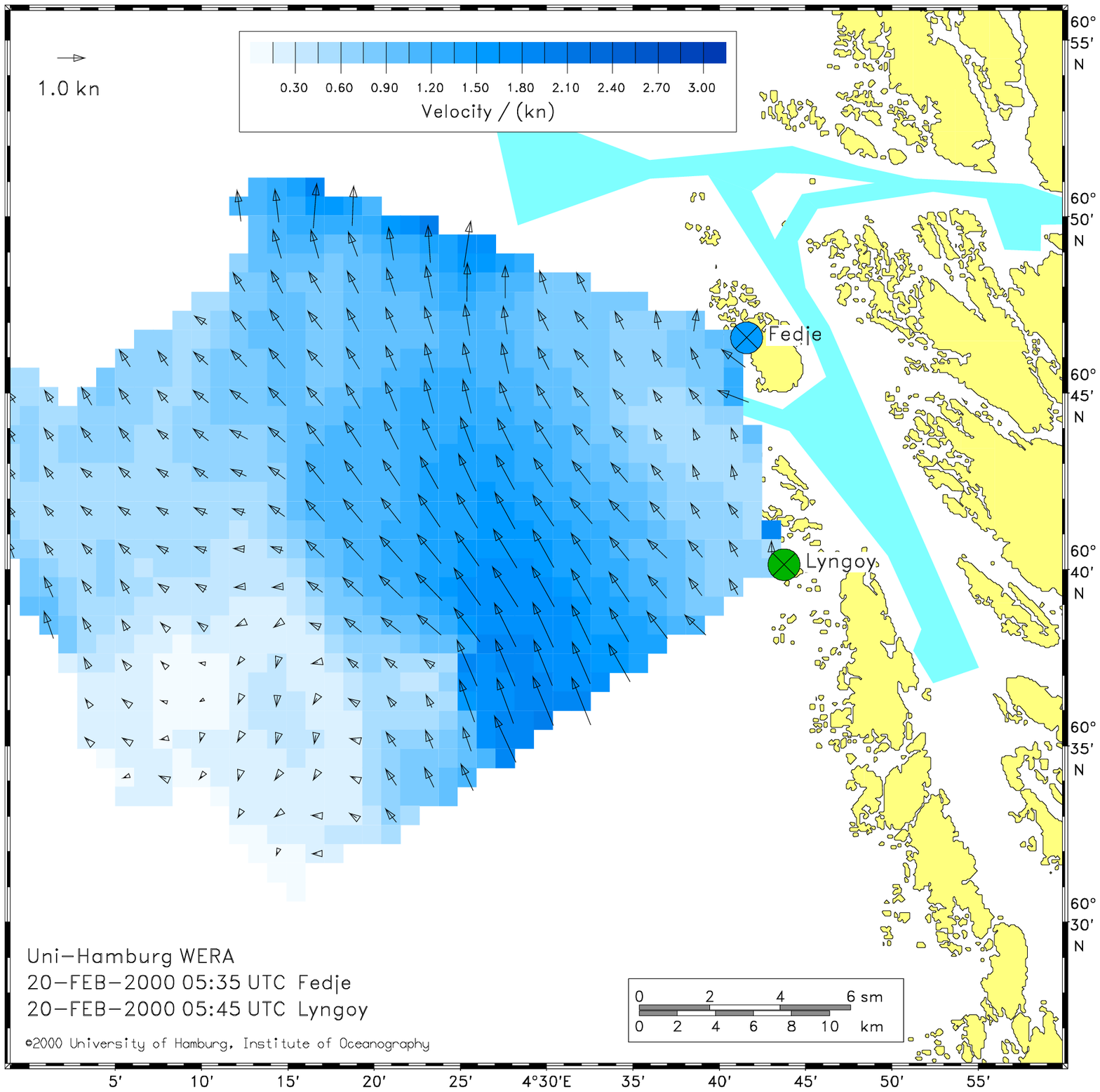, width=0.5\textwidth} \hspace{-0.1cm}
    & 
      \epsfig{file=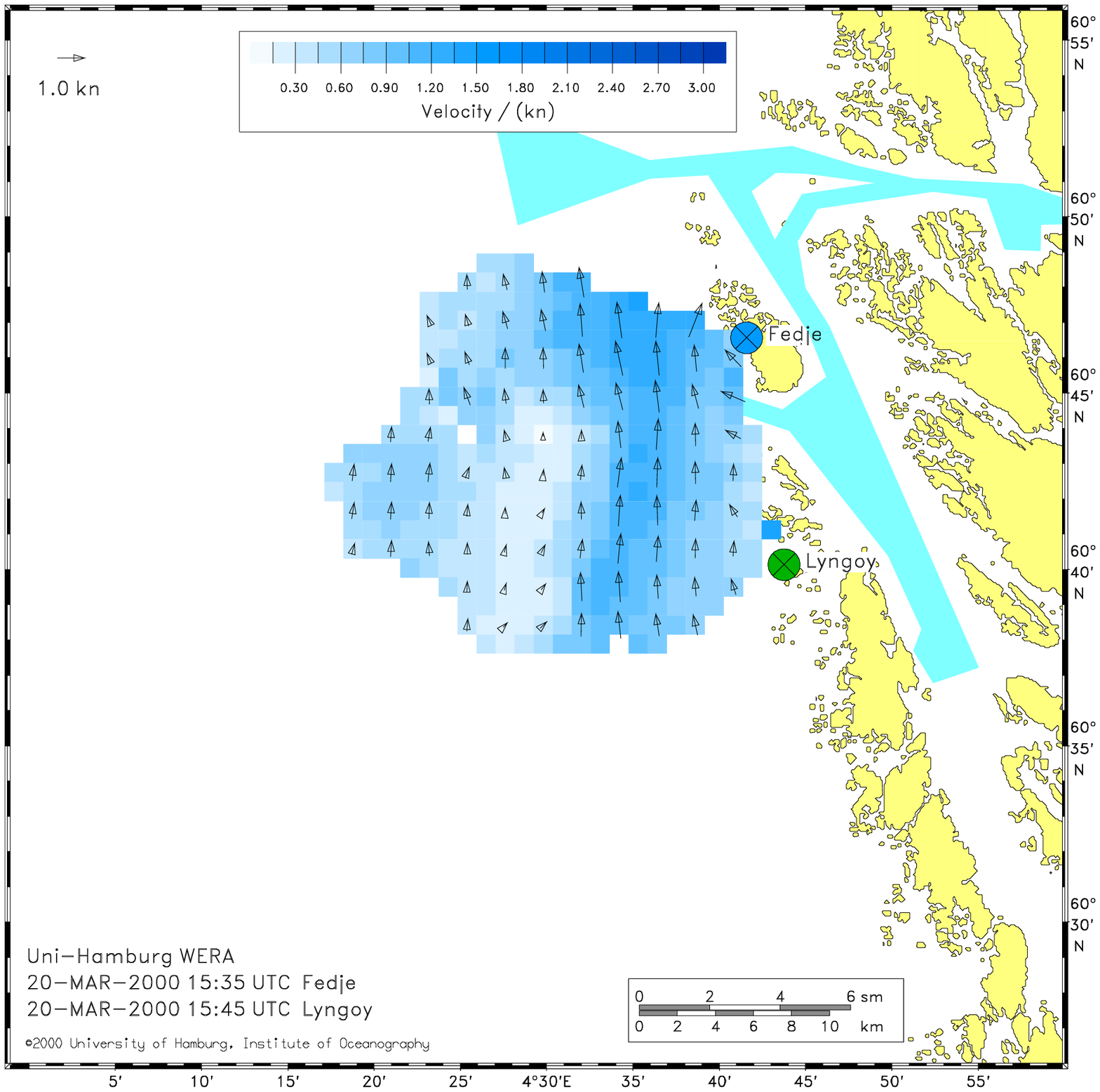, width=0.5\textwidth}
   \end{tabular} }
   \end{center}
 \sfcap{Overview of the Fedje area with typical radar coverage (speed is
 given in knots). The light blue areas mark the ship entrances to the oil
 terminals. The two images illustrate the high variability in radar coverage 
 caused by radio interference, sun spot activity, varying sea state, etc.}
 \label{fig:pryd}
\end{figure}

The first realization of the EuroROSE observing system was focussed on the
island Fedje off the west coast of Norway. This area is bustling with tankers
and other ships that serve the petroleum terminals Sture and Mongstad, together
one of the busiest petroleum harbours in the world. The ship entrances are
narrow and the coastal current is strong and rapidly changing in location and
direction.  \Fig{pryd} provides an overview of the area and typical radar
coverage during the experiment.

As the main objective of the project was to demonstrate the potential of such
an ocean monitoring and forecasting system for coastal security, the
information was transmitted in real time to the local vessel traffic service
(VTS) where observations, analyses, and forecasts were presented automatically.

In this paper, we present the theory behind the assimilation scheme together
with its implementation and performance in the real time monitoring and
forecasting system which was developed for the region around Fedje.
The radar system and the numerical model are also described in some detail.

\section{Sequential data assimilation}
Data assimilation methods have been used routinely by the major weather
prediction centres for the past two decades \cite{dal91}. These methods vary in
complexity from simple Newtonian nudging schemes through optimal interpolation
schemes and all the way up to four dimensional variational assimilation
schemes and sequential Kalman filtering techniques. The latter method has also
been demonstrated for simplified models (the Lorenz equations and quasi
geostrophic models, see \npcite{eve94a,eve97a}).

The fundamental equation in sequential data assimilation is
\begin{equation}
  \psia = \psif + K(\vek{d} - H\psif).
   \label{eq:ass}
\end{equation}
Here, $\vek{\psi} \in \R{n}$ is the $n$-dimensional model state vector, with
superscripts ``a'' and ``f'' denoting analysis and forecast.  Further, $K \in
\R{n\times m}$ is the gain matrix, and $\vek{d} \in \R{m}$ is the data vector
(the $m$ observations).  Finally, $H$ is the observation operator which
maps the model state onto the observation space.

As an explanation to the observation operator, consider a 3D ocean model
yielding horizontal current vectors at certain vertical levels. The data
provided are surface currents measured with a radar facility. Unless the
model and the radar yield data in the exact same locations, some kind of
interpolation is needed to make modelled and observed fields comparable.

The gain $K$ is a matrix of weights that determines each observation's
influence on the final analysis. Using a minimum variance principle (see
\npcite{dal91}), we can argue that $K$ should be
\begin{equation}
  K = P^\mathrm{f}\tr{H}\left(HP^\mathrm{f}\tr{H}+R\right)^{-1},
  \label{eq:gain}
\end{equation}
where $P^\mathrm{f} \in \R{n\times n}$ is the error variance-covariance matrix
of the numerical model, and $R \in \R{m\times m}$ is the error
variance-covariance matrix of the observations.

Let indices $i$ and $j$ denote model variables and $k$ and $l$
denote observations, i.e., $\psif \equiv \{\psi^{\text{f}}_i\}, \;
i = 1,\dots,n$, and $\vek{d} \equiv \{d_k\}, \; k = 1,\dots,m$. The model state
vector mapped to observation space can then be written
$$
   H\vek{\psi} = \{\psi_k\}, \; k = 1, \dots, m,
$$
and the model error variance-covariance matrix is
\begin{equation}
   P^\mathrm{f} = \{\Cov{\psi'_i,\psi'_j}\}, \; i, j = 1,\dots,n.
\end{equation}
Here and throughout, primes indicate zero-mean quantities (deviations from the
mean). 

The gain matrix in \Eq{gain} can be made more understandable by first noting 
that
$$
 P^\mathrm{f}\tr{H} = \{\Cov{\psi'_i,\psi'_k}\}, \; i=1,\dots,n, \, k=1,\dots,m
$$
is simply the error covariance between model variable $i$ and observation $k$.
Secondly, the purpose of the inverse weight matrix in \Eq{gain} is to weight
observations according to their importance. Hence, in a cluster of
observations, one more data point will normally not add much information as the
internal correlation between the observations will be high. Conversely, a
solitary observation in a critical location may be heavily weighted.  The
inverse weight matrix achieves this by balancing the error covariances between
observation locations $k$ and $l$ as predicted by the numerical model,
$$
  HP^\mathrm{f}\tr{H} = \{\Cov{\psi'_k,\psi'_l}\}, \; k,l=1,\dots,m,
$$
against the ``instrumental'' quality of the observations and their internal
covariance, which is contained in the observation error variance-covariance 
matrix,
\begin{equation}
  R = \{\Cov{d'_k,d'_l}\}, \; k,l=1,\dots,m.
\end{equation}

\subsection{Statistical (optimal) interpolation}
Statistical or optimal interpolation (OI) is a sequential data assimilation
method using predefined (time invariant) error statistics.  Hence, the
assimilation (analysis) is variance minimizing \emph{only} if the error
statistics are correct and do not change with time (see \npcite{dal91}).

OI schemes normally assume either that correlations
between two variables $u$ and $v$, say, in different locations are functions 
of distance and direction, 
$$
  \Corr{u'(\xa),v'(\xb)} \approx \text{Corr}_{u',v'}(r,\phi),
$$
or even isotropic functions of distance only,
$$
  \Corr{u'(\xa),v'(\xb)} \approx \text{Corr}_{u',v'}(r).
$$
Here, $r$ is the radial distance $||\xa - \xb||$ and $\phi$ is the direction
from $\xa$ to $\xb$ relative to north.  These simplifications can sometimes be
adequate, but if the covariances display a more complex spatial structure then
this formulation may lead to serious errors in the assimilation update.

\subsection{The Ensemble Kalman filter}
A much more advanced sequential method is the so called Ensemble Kalman filter
(EnKF, see \npcite{eve94a}, \npcite{eve97a}, and \npcite{bur98}). The method is
based on the Kalman filter approach, but circumvents the closure problem of
nonlinearity in the model by calculating the covariances from an ensemble of
evolving model states. This approach also avoids forecasting the $n\times n$
covariance matrix $P^\mathrm{f}$ which becomes intractable for 3D oceanic and
atmospheric models of realistic dimensions.

An ensemble of $N$ initial model states 
$$
   \{\vek{\psi}_\nu^0\}, \; \nu = 1, \dots, N,
$$ 
is generated and integrated forward in time, yielding an ensemble of forecasts
at a later time $t_1$,
$$
   \{\psif_\nu\}, \; \nu = 1, \dots, N.
$$ 
The ensemble average is 
\begin{equation}
   \psm = \frac{1}{N} \sum_{\nu=1}^N \bmpsi_\nu,
\end{equation}
and the deviations from the mean are denoted $\bmpsi_\nu' \equiv \bmpsi_\nu -
\psm$.

The zero mean ensemble can also be viewed as a collection of column vectors,
illustrated by the following tableau,
\begin{equation}
  \bvec{& A' &} = 
   \begin{bmatrix}
      \bvec{\bmpsi'_1}, & \dots, & \bvec{\bmpsi'_N}
   \end{bmatrix}.
\end{equation}
The model error covariance matrix can be approximated by the outer product of
the ensemble,
\begin{equation}
   P^\mathrm{f} \approx \frac{1}{N-1}A'\tr{A'}.
   \label{eq:P}
\end{equation}
At each update, the Kalman gain $K$ must be found from \Eq{gain} by first
integrating $N$ numerical model realizations (where $N$ is typically
$\mathcal{O}(250)$) and then computing $P^\mathrm{f}$ using \Eq{P}.  Clearly,
this method is computationally very expensive, and it would be desirable to
find a middle ground between the over-simplification of standard OI and the
enormous cost of running $N$ sibling models.

\subsection{A ``quasi-ensemble'' assimilation scheme}

\label{sec:quasi}
We propose to exchange the ensemble of models used to compute the Ensemble
Kalman Filter with an ensemble of \emph{model states}. This can be any
collection of model states taken from a representative model run. Whereas an
ensemble of models will pick up periods of high and low variance,
our covariances will remain fixed throughout the assimilation period. This
means that an assimilation scheme employing these statistics will formally
belong to the class of previously discussed optimal interpolation schemes.

In order to generate this quasi ensemble, we need to run the model for a
representative period. Model states are sampled from the model run.  This
reference period should be selected carefully. Ideally, it should pick up the
correlations relevant for our specific time period. However, this may be
impossible to do in advance, as is the case for a real time observing system.
For a \emph{hindcast} experiment, one should naturally choose the reference run
to cover the period of interest. The next best thing will be to choose a
reference run that covers a similar period (similar climatology, e.g., a
different year but same time of year).

A final caveat regarding the sampling is to choose a sampling frequency that
captures the relevant physics. This is especially critical for the tidal motion
with its well-defined frequencies. This is most easily achieved by choosing a
sampling period $\Delta t$ that is smaller than the dominant tidal period, but
\emph{not} an integer fraction of it (to avoid capturing only high tides or only
low tides).

An EnKF will find the ``correct'' error statistics under the assumptions that
the ensemble is sufficiently large and the errors are normally distributed. The
proposed quasi ensemble will not, in general, find the correct error variances
for the model since its statistics do not vary with time. However, the
\emph{correlations} may be satisfactory. This means that although the relative
weight between data and model may be off, the spatial correlations between, say,
surface currents in one point and the salinity in another point, may be quite
all right.

The formulation of the quasi-ensemble filter is analogous to the derivation of
the EnKF and makes no assumptions on the spatial shape of the covariances. In
theory, cross-correlations between all pertinent model variables may be used to
make full use of the data, hence model grid points well outside the area
covered by observations may experience corrections based on their covariance
with variables in points closer to the observations. Likewise, vertical
corrections can be made to the different model variables. The full equation
reads
\begin{equation}
 \psia = \psif +
 \frac{1}{N-1}A'\tr{A'}\tr{H}\left(\frac{1}{N-1}HA'\tr{A'}\tr{H}+R\right)^{-1}
 \left(\vek{d}-H\psif\right).
 \label{eq:quasi}
\end{equation}
Here, we have substituted the ensemble (of sampled model states) for
$P^\mathrm{f}$ using \Eq{P}.

\section{The EuroROSE current assimilation and forecasting system}

\subsection{The numerical model} 
\label{sec:model}
We used the Princeton Ocean Model (POM) as implemented and modified by The
Norwegian Meteorological Institute (DNMI).  The lateral hydrodynamic equations
are solved on an Arakawa C-grid (see \npcite{mes76}, \npcite{kow93} p~170).
Terrain following $\sigma$-coordinates resolve the vertical, which means that
the vertical resolution is high in shallow areas and coarse in deeper areas. In
addition, the levels are not distributed linearly over the depth as the mixed
layer near the surface requires higher resolution than the deeper parts of the
ocean. Surface elevation and the vertical component of the current vector, $w$,
are solved on the surface. Through the water column, $u$ and $v$ are staggered
one half grid cell with respect to $w$. Hence, the horizontal velocities of the
uppermost layer are found one half grid cell from the surface.

The model contains a second-order moment turbulence closure sub-model
\cite{mel82} which provides vertical mixing coefficients.  The model solves
the conservation equations for momentum and mass using an explicit finite
difference scheme in the horizontal, and an implicit scheme for the vertical
terms to eliminate the time-step constraints caused by fine resolution of the
surface layer. The model has a free surface and uses mode-splitting for the
time-stepping. In the external mode, the model is two-dimensional and uses a
time-step limited by the Courant-Friedrichs-Lewy (CFL) criterion for fast
propagating barotropic waves.  For the internal mode, the model is
three-dimensional and uses a longer time-step based on the CFL-criterion for
internal wave speed. A Leapfrog scheme is used for the advection terms.
The model version developed at DNMI has a Flow Relaxation Scheme (FRS)
implemented at the lateral open boundaries \cite{mar87}, where forcing from
eight tidal constituents is also included \cite{eng95}.
A thorough description of the basic model setup can be found in
\incite{blu87}. For further information on the modifications made to the DNMI
version of the model please refer to \incite{eng95}.

\subsubsection{The model realization}
\begin{figure}[t!]
 \centerline{
  \epsfig{file=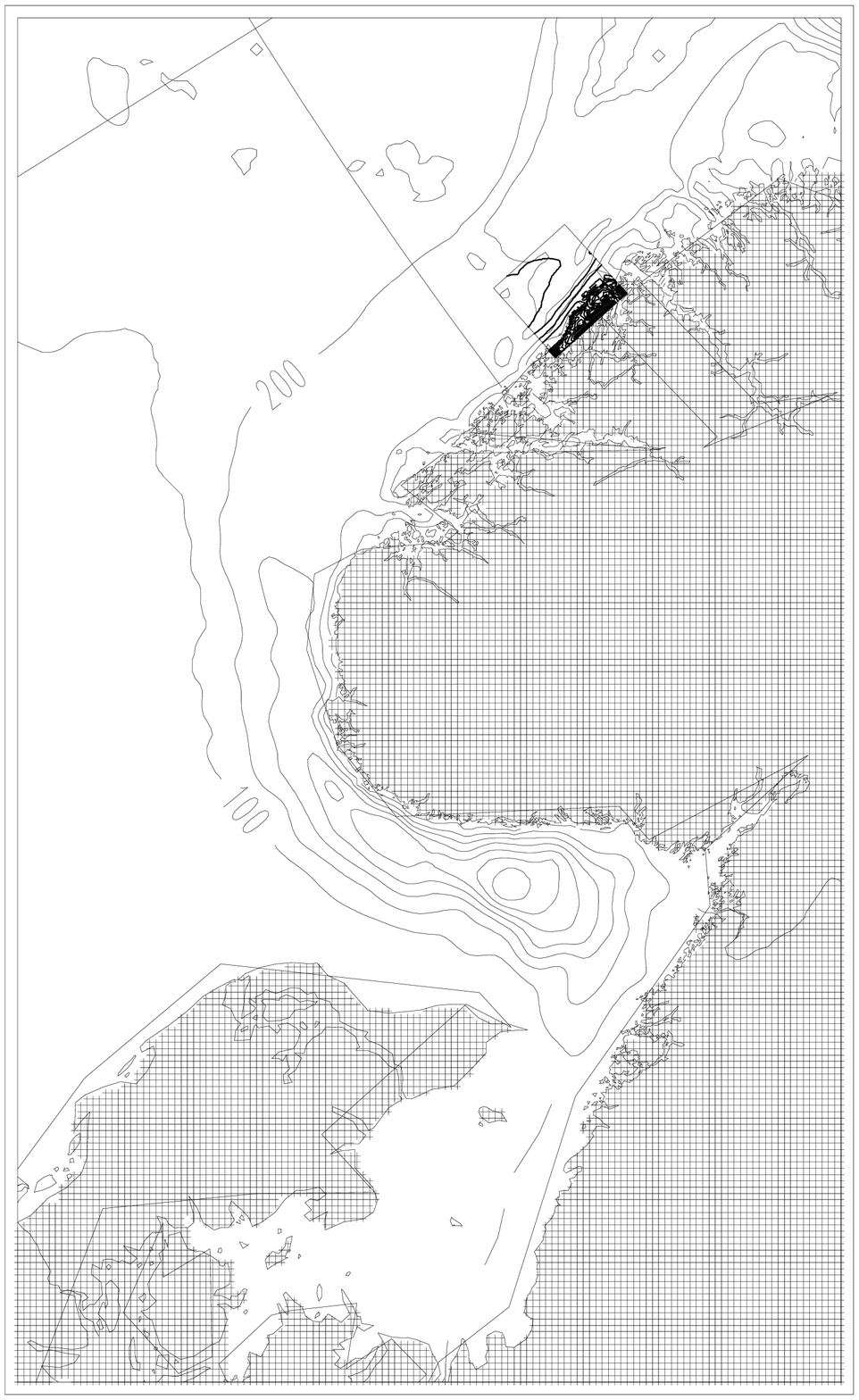, width=0.7\textwidth}
 }
 \sfcap{The 1~km high resolution model domain is shown as a small square
 superposed on the bathymetry of the 4~km intermediate model covering parts of
 the North Sea, Skagerrak, Kattegat, and the coastal waters around southern
 Norway. The projection is polar stereographic, north is toward the upper right
 hand corner of the map.}
 \label{fig:neurope}
\end{figure}

\begin{figure}[h]
 \centerline{
  \epsfig{file=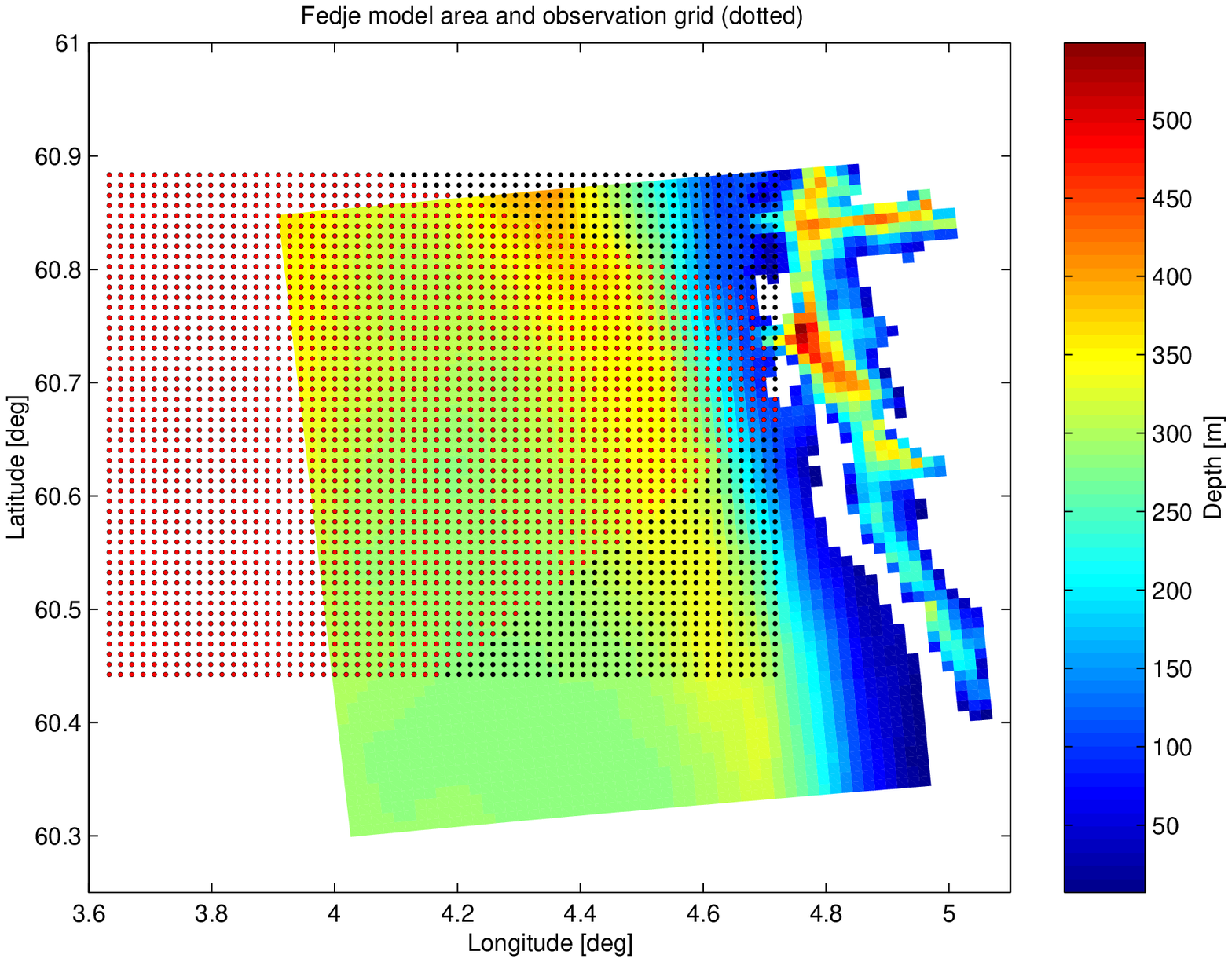, width=0.8\textwidth}
 }
 \sfcap{The HF radar grid (dotted) superposed on the topography of the inner
 model. Inshore of the two islands, the topography has been manually corrected
 using ship charts to permit more detail around the ship entrances.}
 \label{fig:obsgrid}
\end{figure}

Three models are nested inside each other. The outer model covers
the North Atlantic and the Norwegian sea with a resolution of approximately
20~km. The intermediate model covers the coastal waters of southern Norway
with a resolution of 4~km (see \Fig{neurope}), and the inner, high resolution
model has a resolution of 1~km and covers a $60\times 60$~km area (see
\Fig{obsgrid}). The inner model has 17 $\sigma$-levels which are distributed as
follows (values are given as parts per thousand of total depth),
\begin{equation}
  \sigma_k = \tr{[0,2,5,10,25,50,100,150,200,300,400,500,600,700,800,950,1000]}.
  \label{eq:sigma}
\end{equation}
The above distribution of $\sigma$-layers is illustrated in \Fig{zcorruv} with
symbols ``o'' and ``x'' on the correlation profiles.

The main part of the area covered by the radar antennae has a water depth of
$\sim300$~m. At this depth, the height of the uppermost grid cell is $\sim
0.6$~m (cf \Eq{sigma}). As the horizontal current components are staggered with
respect to the vertical component, the uppermost horizontal currents are
computed at a depth of $\sim0.3$~m.  Measurements made with HF radar are
essentially weighted averages of the currents in the vertical column that
``feel'' the ocean Bragg wave \cite{fer96}. The radar frequency is $27.65$
MHz. Its backscatter is in resonance with an ocean wave of $\sim5$~m. The
corresponding average current depth is estimated to be $\sim0.5$~m, hence the
radar currents are comparable to the currents in the uppermost model layer.

The models are matched using a flow relaxation zone extending seven grid points
into the model domain. All three models are run on rotated polar stereographic
grids.  The topography used in the high resolution model is taken from the
\texttt{ETOPO5} database which is publicly available on the Internet.  The
dataset is rather too coarse for our application with a resolution of
approximately 4.5~km east-west and 9~km north-south. In order to get the
entrances between the islands right, we used ship charts to manually correct
the topography of the inner model. This explains why \Fig{obsgrid} exhibits
rather more detail in the estuary than along the coast. Tides are included in
the intermediate model and are propagated as a barotropic signal into the
nested model. No ice model is included, as we are in a permanently ice-free
part of the Norwegian Sea. Monthly mean values for river runoff are included in
all three models. The wind fields and boundary values are updated every three
hours. The outer models are forced with 50~km resolution winds from a limited
area atmospheric model (LAM). The inner model is forced using winds from a
10~km resolution atmospheric model to include topographic effects near the
coast. Both atmospheric models and the two outer ocean models are run routinely
by DNMI.


\subsubsection{The model error statistics}
As explained in \Sec{quasi}, the assimilation scheme requires a reference run to
compute the model error covariance matrix.  In our case, the model
period was Feb-Mar 2000, and consequently we ran the model for the same period
in 1999 with a sampling period $\Delta t = 5.5$~hours, assuming that ``on the
average'' this model run would pick up the important correlations inherent to
the model state. The choice of $\Delta t$ is guided by the observations made
in \Sec{quasi} that integer multiples of the major tidal constituents
should be avoided. A period of 5.5~h will march slowly through the tidal cycle
and thus avoid sampling only, say, the high tide and the ebb.

The result of this reference run was then scrutinized and found
to yield sensible results. The full correlation matrix of the model can be seen
as a six-dimensional object,
\begin{equation}
   \text{Corr}(\psi'(\vek{x}_1),\psi'(\vek{x}_2)),
\end{equation}
where $\psi'(\vek{x}_1)$ and $\psi'(\vek{x}_2)$ may assume all model variables
in all different grid points in the model domain.

\begin{figure}[h]
 \begin{center}
 \centerline{
 \begin{tabular}{ll}
  \epsfig{file=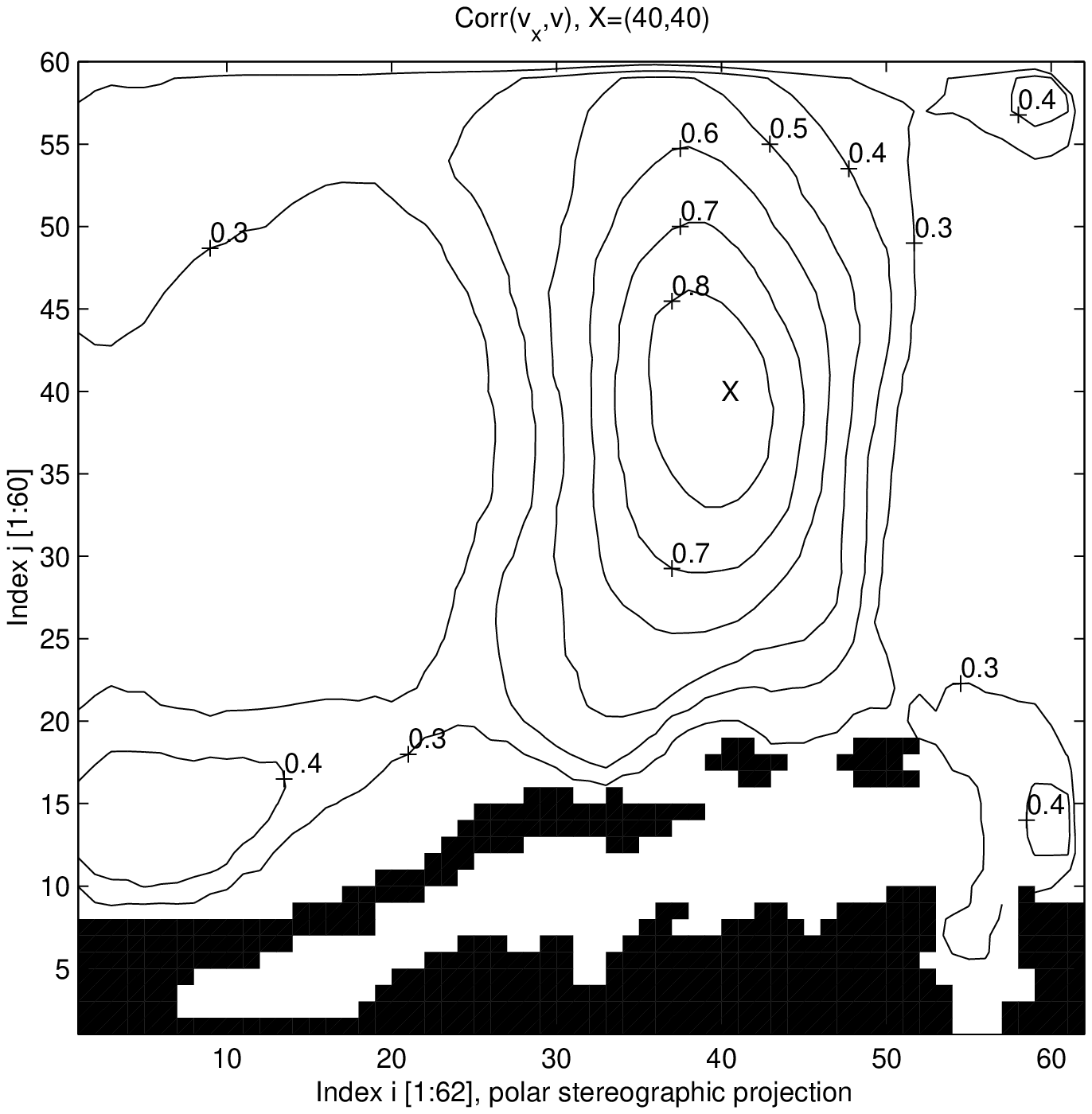, width=0.5\textwidth} \hspace{-0.5cm} &
  \epsfig{file=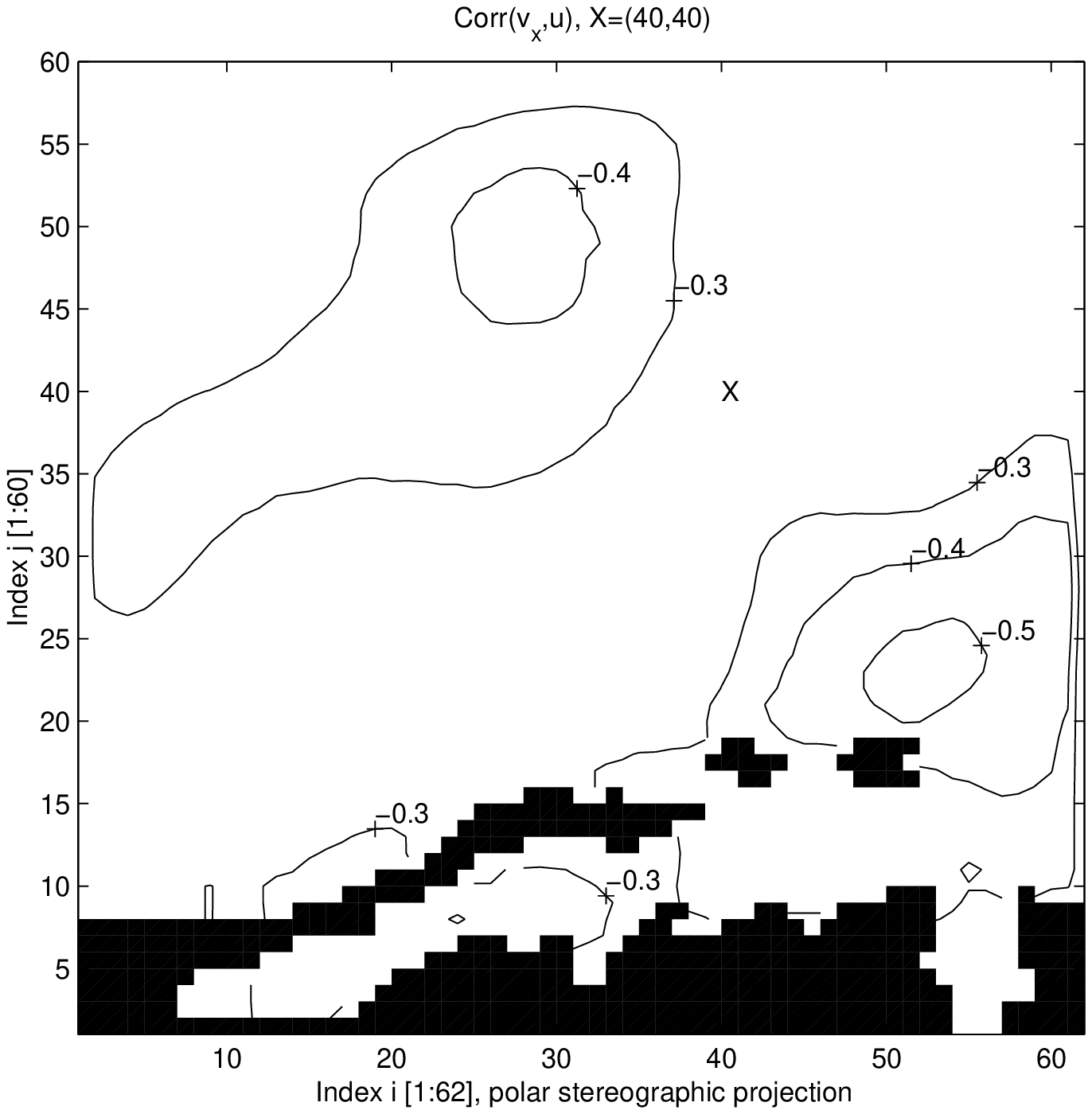, width=0.5\textwidth}
 \end{tabular} }
 \end{center}
 \sfcap{The horizontal correlation between $v$ (left panel), $u$ (right panel)
  and $v$ in point X(40,40,1). The correlation plots have been smoothed using a
  $6\times6$ median filter and given a treshold value of $\pm 0.3$ to hide
  weak, irrelevant correlations. The projection is polar stereographic. The
  model land mask is superposed for geographic reference.}
 \label{fig:vcorrvu}
\end{figure}

In \Fig{vcorrvu}, a slice through the model correlation matrix has been made by
freezing $\vek{x}_2$ at grid point (40,40,1), marked with an ``X''. This point
was chosen because it lies in the centre of the radar coverage. The model grid
orientation is such that $u$ is the alongshore current and $v$ the across-shore
current. The across-shore current correlation shows the expected behaviour of
decorrelation with radial distance. Further, the cross correlation between $u$
and $v$ is found to be quite strong. Its maximum is shifted away from the
location X, meaning that the maximum cross correlation does not occur on the
spot.

\begin{figure}[h]
 \begin{center}
 \centerline{
 \begin{tabular}{ll}
  \epsfig{file=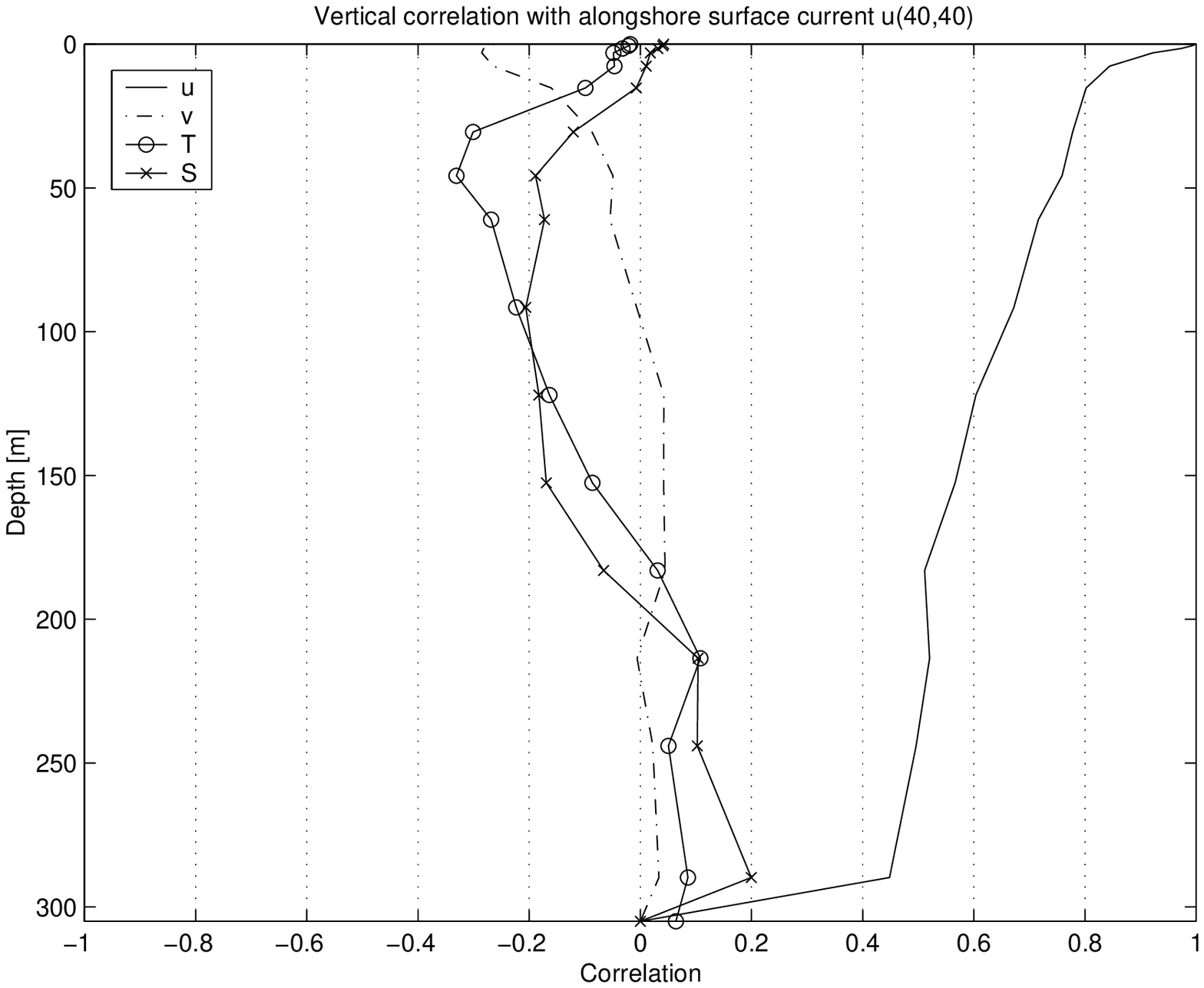, width=0.5\textwidth} \hspace{-0.5cm} &
  \epsfig{file=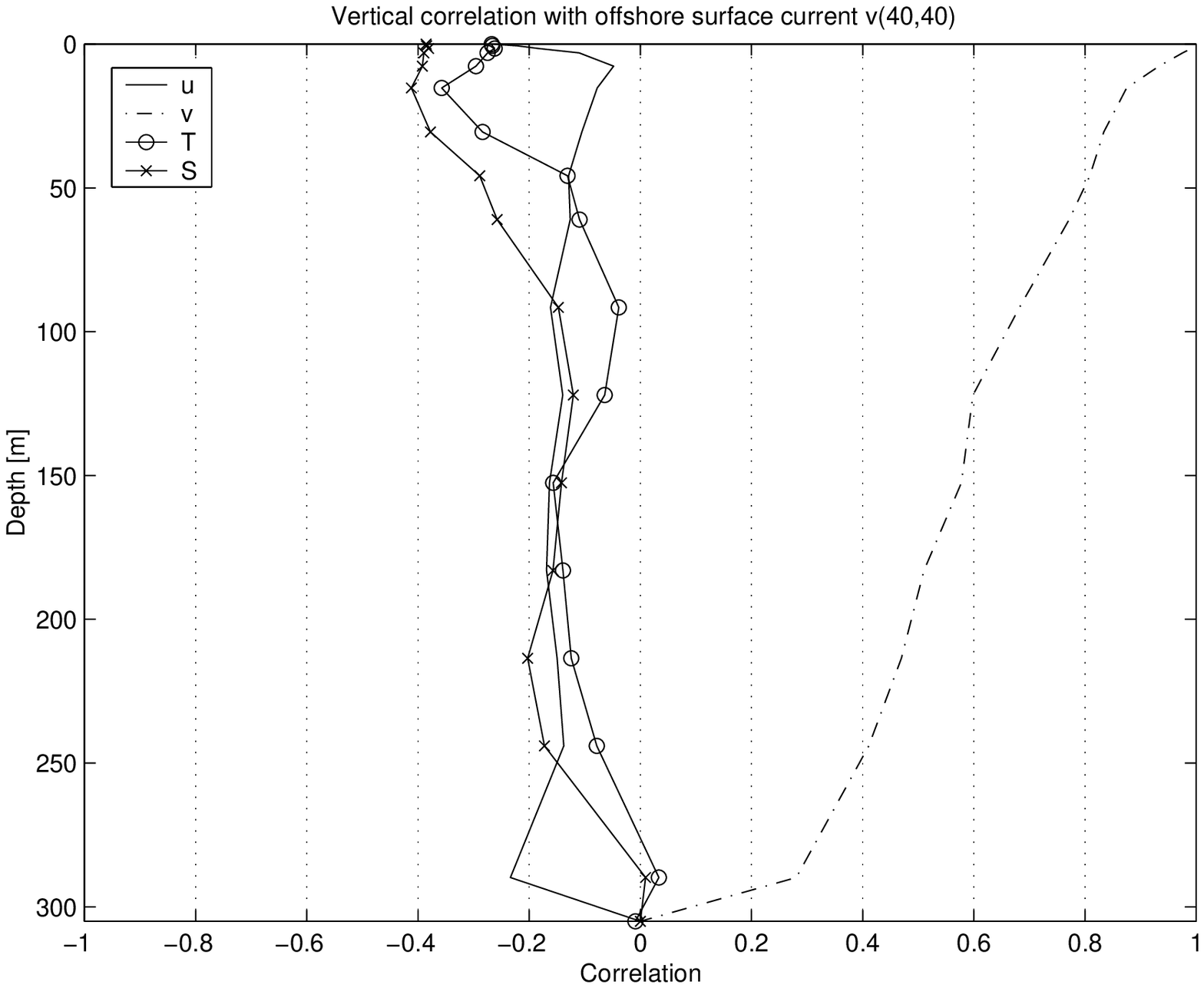, width=0.5\textwidth}
 \end{tabular} }
 \end{center}
 \sfcap{The vertical cross correlation between $u$ (left panel), $v$ (right
 panel) and variables $u$, $v$, $T$, and $S$ in point X(40,40,1).}
 \label{fig:zcorruv}
\end{figure}

The vertical correlation structure with the surface current in point X(40,40,1)
is shown in \Fig{zcorruv}. The model layer depths are indicated with symbols
``o'' and ``x'' in the cross correlation with $T$ and $S$. As is seen, the
model resolves very well the upper 50 m of the ocean. This is necessary to
capture the influence of the atmosphere on the oceanic mixed layer. In general,
the surface currents do not correlate strongly with the hydrography of the
underlying water masses. The strongest correlations are of the order of -0.4.

The across-shore current exhibits a slightly stronger correlation with the
hydrography than the alongshore current. Negative correlation indicates that
across-shore currents advect fresh and cold coastal waters away from the coast
(remember that the location of the point X(40,40) is quite far from shore).
The alongshore surface current $u$ is completely detached from
the hydrography and only at about 50~m depth does the correlation rise above an
absolute value of 0.2.

The model statistics reveal a strong surface current to deeper current
correlation, but a significantly weaker cross correlation between current
components.  Note also the immediate drop in correlation at the surface to
about 0.8 at 20 m depth. This illustrates how the wind energy is distributed in
the upper water column.

\subsection{The data}
Our observations are taken from a beam forming phased array HF radar called
``Wellen Radar'' (WERA). The radar is operated by the University of Hamburg and
was temporarily mounted on the islands Lyng{\o}y and Fedje.  Each radar array
measures the radial component of the surface current. For a walk through the
theoretical underpinnings of the current measurements, refer to
\incite{ess00}, \incite{gur97} and \incite{gur99}. The radar range is affected
by the sea state, atmospheric disturbances, radio interference and sun spot
activity. \Fig{pryd} illustrates the variability in radar coverage.

The data are delivered on a regular grid in geographical coordinates (longitude
and latitude, see \Fig{obsgrid}). The resolution of this grid is 1~km, which
makes the data density comparable to the model grid, albeit with different
orientation.  Independent measurements of both $u$ and $v$ can only be found in
a triangle covered by both arrays (see \Fig{obsgrid}).

\begin{figure}[h]
 \centerline{
  \epsfig{file=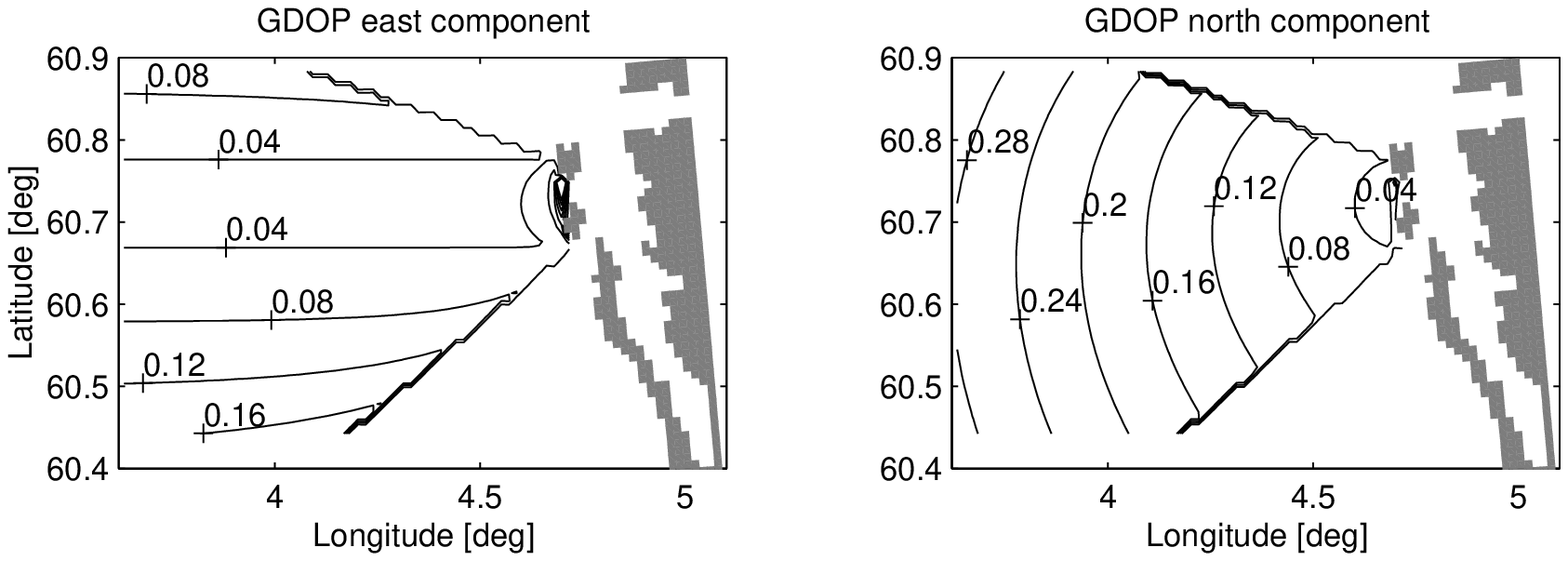, width=1.0\textwidth}
 }
 \sfcap{The geometric dilution of precision (GDOP) in the east component (left
 panel) and the north component (right panel) of the HF radar current vectors.
 The model land mask is shown for geographic reference.}
 \label{fig:gdop}
\end{figure}

The geometric dilution of precision (GDOP) is the geometric error made when
combining two radial current components. This error reaches a maximum when the
two radial components are nearly in the same or in opposite directions, see
\incite{cha97}. The GDOP has been computed for the east and north components of
the radar current vectors (see \Fig{gdop}). These time-invariant error
variances enter the diagonal of the error variance-covariance matrix $R$ in
\Eq{gain} to weight observations against the forecast. No information is
available on the error covariance between observations, hence $R$ is assumed to
be diagonal. We have chosen to only include the time-invariant part of the
observation error (assuming an RMS error of 5 cm/s on the radial components and
then computing the GDOP) to speed up the observing system. To compensate for
this procedure, we have included an extra data quality check described in
\Sec{quality}.

\subsection{The assimilation cycle}
\begin{figure}[h]
 \centerline{
  \epsfig{file=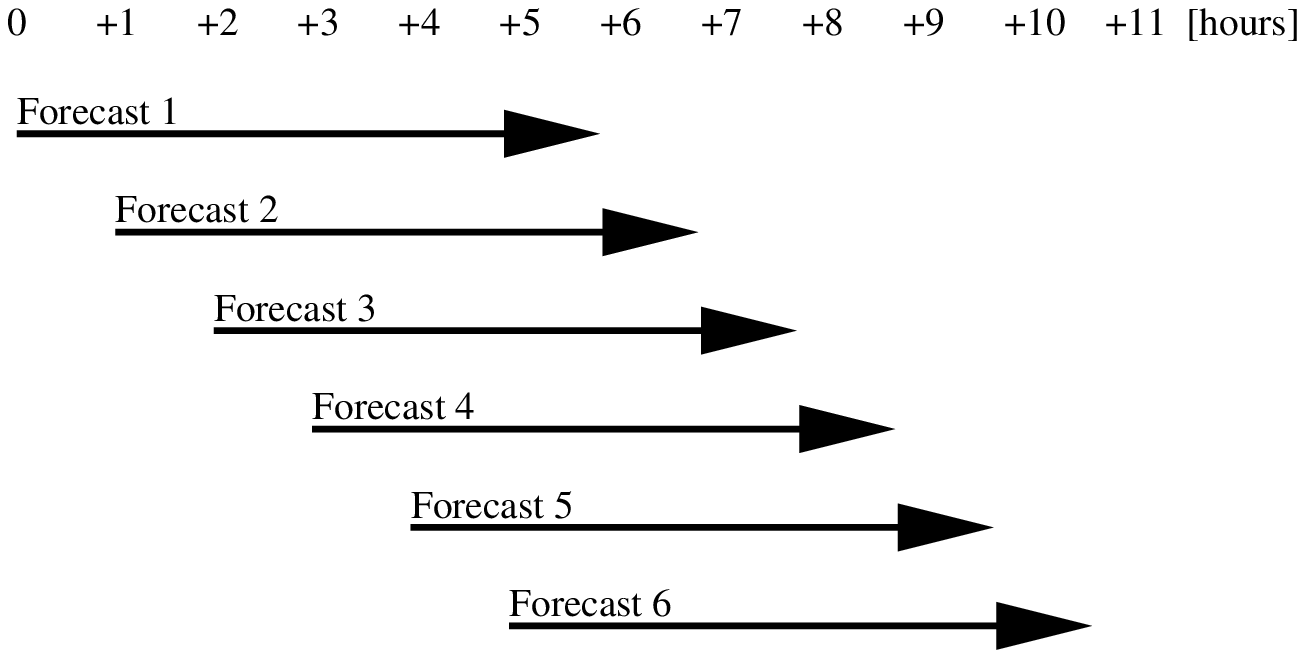, width=0.5\textwidth}
 }
 \sfcap{The analysis and forecast cycle. A new assimilation and six hour
 forecast is started every hour on the hour around the clock. The analysis and
 forecasts are automatically transferred to the Vessel Traffic Service at Fedje 
 for immediate presentation.}
 \label{fig:cycle}
\end{figure}

New data arrive every 20 minutes from the radar. Each model cycle consists of
assimilation at times -00:40, -00:20, and 00:00 relative to the analysis time.
After this, the model generates a six hour forecast (see Table~\ref{tab:cycle}
and \Fig{cycle}). This cycle is repeated every hour, resulting in several
overlapping forecasts. Because we assimilate every 20 minutes, the initial
field that starts the next assimilation cycle is a 20 minute forecast based on
the assimilation in the previous cycle.
\begin{table}[h]
\centerline{
\begin{tabular}{|r|l|} \hline
Time [h] & Action \\
\hline \hline
-1:00 & Model is initialized from previous assimilation cycle \\
\hline
-0:40 & First dataset in, assimilation \\
\hline
-0:20 & Second dataset in, assimilation \\
\hline
0:00  & Third dataset in, final assimilation \\
\hline
0:00  & Time of analysis, forecast begins \\
\hline
$\vdots$ & $\vdots$ \\
\hline
6:00 & Forecast ends \\
\hline
\end{tabular}
}
\sfcap{The analysis and forecast cycle. This is repeated every hour 
throughout the period yielding six overlapping forecasts and one 
analysis at anyone time. Times are given as lead times relative to analysis
time.}
\label{tab:cycle}
\end{table}

\begin{figure}[h]
 \centerline{
  \epsfig{file=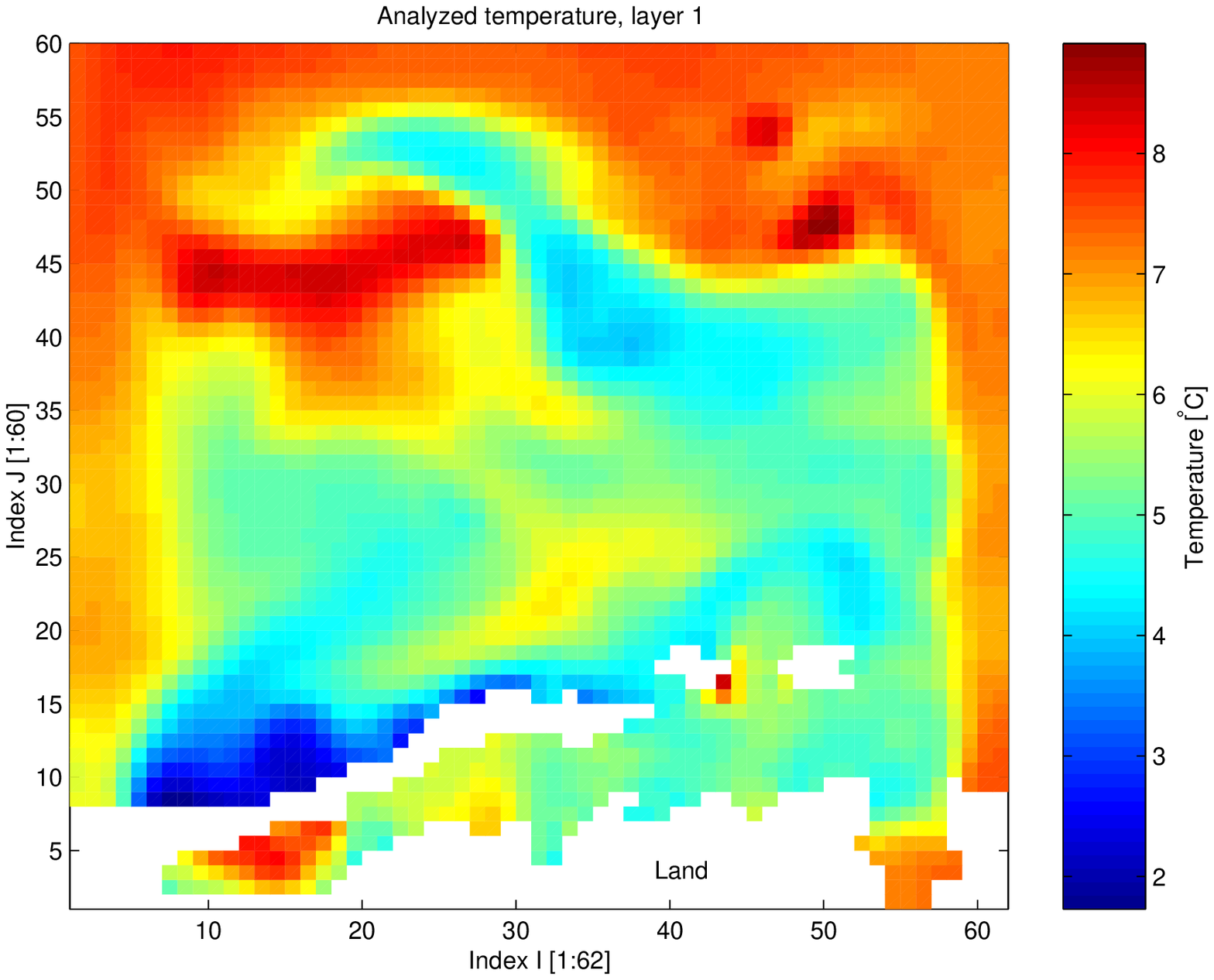, width=0.8\textwidth}
 }
 \sfcap{The temperature field after correcting $T$ and $S$
 with HF radar surface current observations. The outer model is oblivious of
 the update of the inner model, and hence gradients in temperature and salinity
 build up along the open boundary, eventually causing the model to go
 unstable.}
 \label{fig:bad}
\end{figure}

At the outset, it was considered worthwhile to include the cross-correlations
between surface currents and hydrography ($T$ and $S$, see \Fig{zcorruv}), both
horizontally and vertically. This way the density field could be corrected to
make it consistent with the observed velocity field. However, even after
smoothing the analysis using a second order Shapiro filter (see \Sec{shapiro}),
we found the model to go unstable. It turns out that because the model is
nested inside an external model that is oblivious of any assimilation,
corrections in $T$ and $S$ cause a density gradient to build up along the open
boundary. This sets up an unphysical circulation which eventually blows up the
model (see \Fig{bad} for an example). To save the assimilation experiment, we
were forced to leave out cross-updates of hydrography. This limits the memory
of the assimilation, as the hydrography is no longer consistently updated with
the current field and will cause the current field to lapse back to its
original state more quickly.  The only solution to the observed mismatch
between our nested models is to perform an assimilation in both models (inner
and outer). That, however, would be a much more extensive task and was not
considered possible in a real time experiment like this. In light of the weak
correlations found between surface currents and hydrography (\Fig{zcorruv}) we
conclude that in this model realization it is better to leave out corrections
to $T$ and $S$ anyway.

\subsubsection{Data quality control and smoothing}
\label{sec:shapiro}
\label{sec:quality}
To avoid extreme updates in the model due to bad data, a comparison between the
modelled and observed currents is made for each observation prior to analysis.
If the observed speed differs by more than $0.5$ m/s, or the observed direction
deviates by more than $45^\circ$ from the modelled current vector, the
observation is discarded. These thresholds have been chosen rather arbitrarily,
but have stood the test and proved to weed out the strong, erroneous current
vectors that are often found along the rim of the radar maps (due to
backscatter from the strong antenna pattern sidelobes found on the edges of the
radar coverage). This quality check is also a way to compensate for the lack of
time-varying observation error variances. On the average, only a small
percentage of the observations were discarded through this procedure (less than
5\%). However, in situations where the radar performed poorly, larger amounts
of data were discarded.

Further, the assimilation sometimes adds too much fine structure to the model
fields. To remove this but retain the longer wavelengths we chose to run a
Shapiro second order nine-point filter after each analysis. It has been shown
(see \npcite{sha70} or \npcite{hal80}) that the Shapiro filter removes the
``$2\Delta x$'' wave completely (the shortest wave that can be represented,
dictated by the Nyquist frequency), while the longer wavelengths are not much
dampened (zero damping for infinite wavelength). The ``$10\Delta x$'' wave is
attenuated by less than 10\%.

\subsubsection{Restarting the model}
\label{sec:restart}
The ocean model uses centered differences in space and time (leapfrog scheme)
for the integration of the horizontal momentum equations.

As a demonstration of the scheme, we take the one-dimensional advection
equation
\begin{equation}
   \Dp{u}{t} + c\Dp{u}{x} = 0
   \label{eq:advect}
\end{equation}
where $c$ is the (constant) advection velocity. Applied to this equation, the
leapfrog scheme becomes
\begin{equation}
   u^{n+1}_i = u^{n-1}_i + c\frac{\Delta t}{\Delta x} \left(u^n_{i+1} - 
     u^n_{i-1}\right).
   \label{eq:leapfrog}
\end{equation}
Here, $i$ is the spatial index and $n$ the temporal index. Further, $\Delta
x$ and $\Delta t$ represent the spatial and temporal resolution, respectively.

The assimilation scheme is invoked at time $t_n$, which is incompatible with
the above scheme because it involves $t_{n-1}$, and hence the corrections
introduced by the assimilation would not be propagated forward in time (as
the scheme ``leapfrogs'' past time $n$ from time $n-1$ to $n+1$). To circumvent
this, we have to apply a one level scheme immediately after the update,
\begin{equation}
   u^{n+1}_i = u^n_i + c\frac{\Delta t}{2\Delta x} \left(u^n_{i+1} - u^n_{i-1}
     \right).
   \label{eq:euler}
\end{equation}
This is known as an Euler scheme and is unconditionally unstable for all
wavelengths \cite{hal80}. However, the scheme can still be used for a few
timesteps at a time as long as we revert to a stable scheme later.

\section{Assimilation performance}
The observing system was active for a period of approximately six weeks. The
assimilation scheme in its final form was used for three weeks. Throughout the
experiment, an identical model twin was run without assimilation (hereafter
referred to as the free run). This freerunning model allows us to assess the
impact of the assimilation scheme on analysis and forecasts. In the following,
we have compared the two model runs with the radar data in wont of other
sources of ground truth. It is important to keep in mind that the radar data
themselves have errors.

\begin{figure}[h]
   \begin{center}
   \centerline{
   \begin{tabular}{ll}
      \begin{sideways}
      \epsfig{file=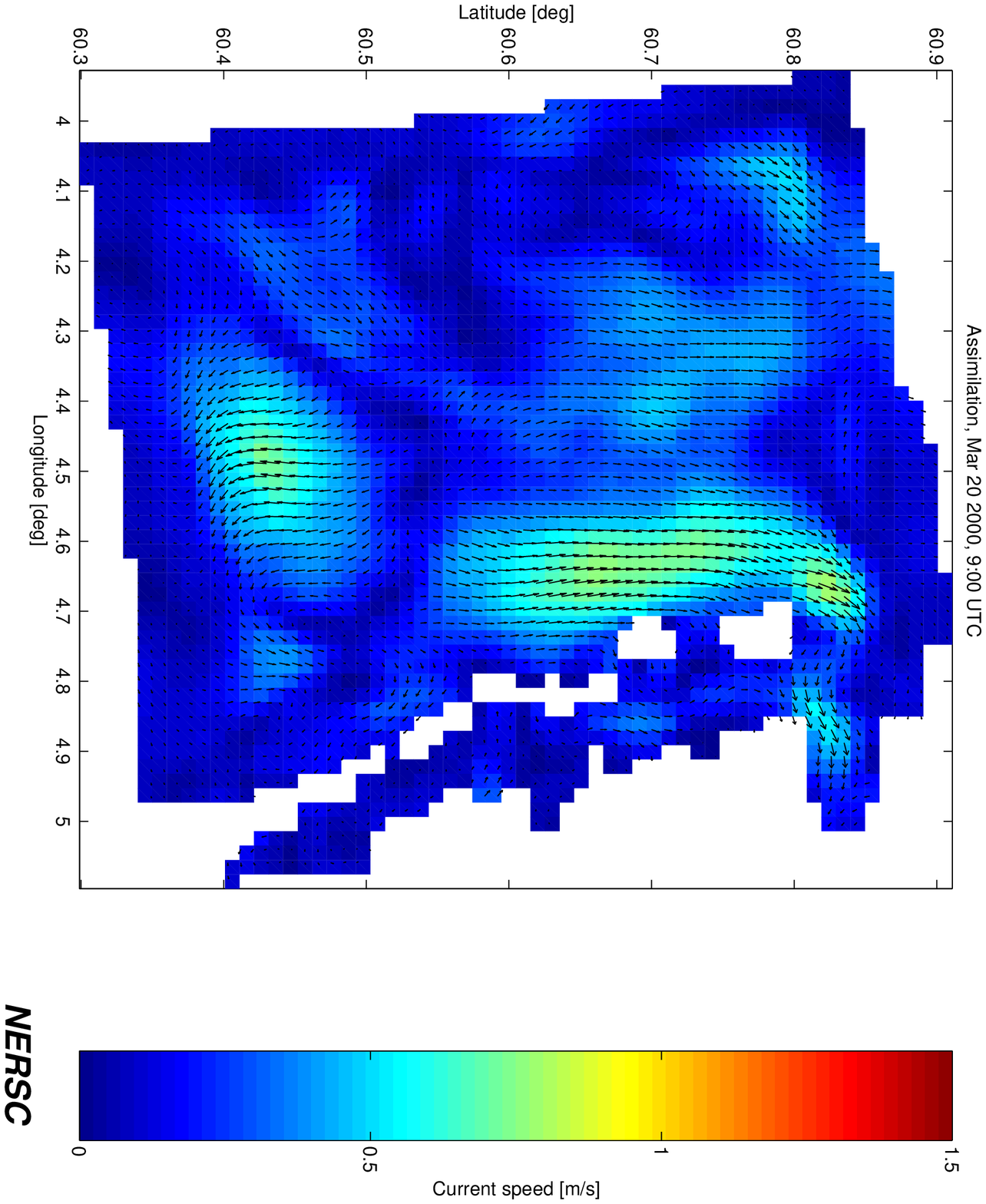, width=0.5\textwidth} \hspace{-0.7cm}
      \end{sideways}
    & \begin{sideways}
      \epsfig{file=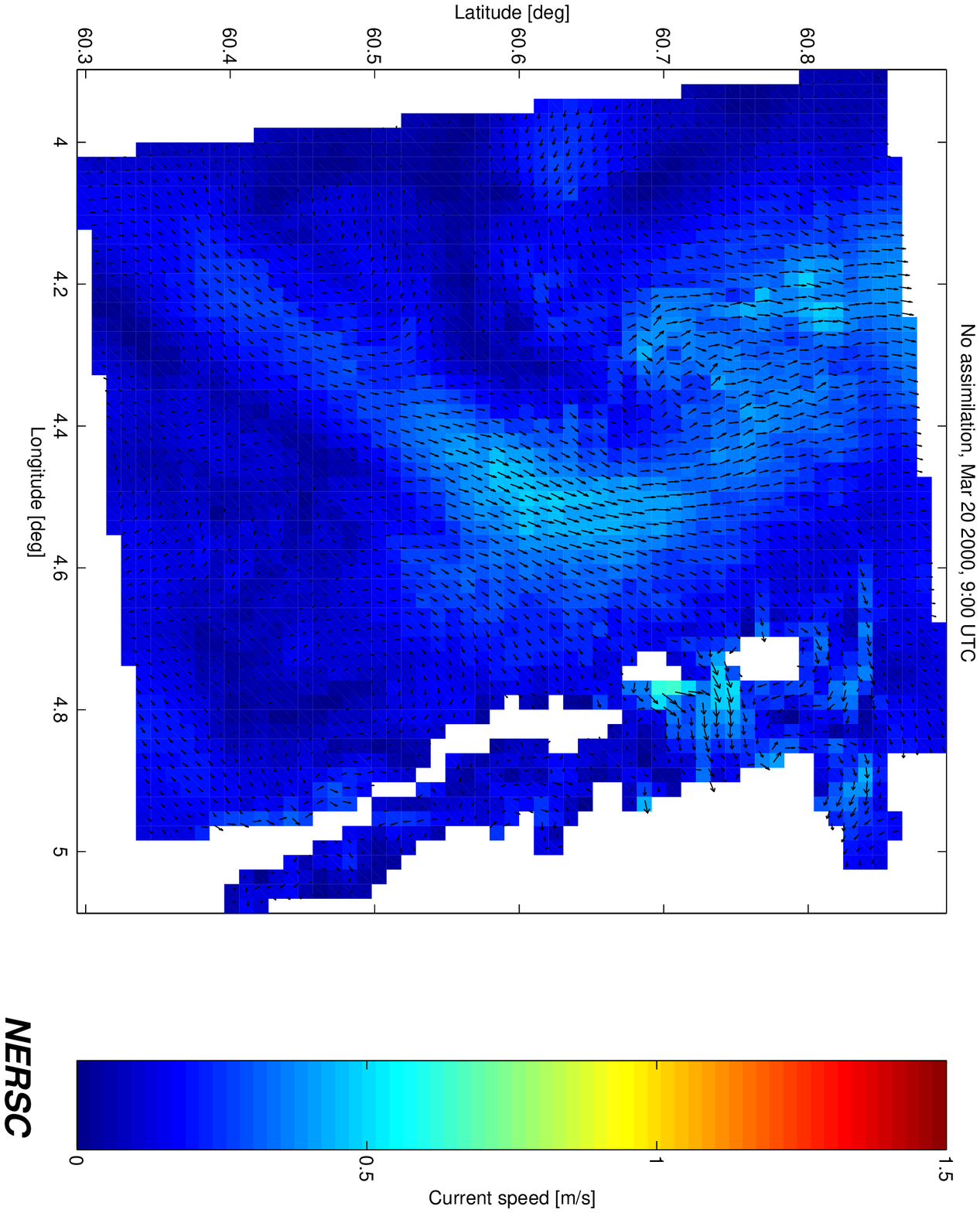, width=0.5\textwidth}
      \end{sideways}
   \end{tabular} }
   \end{center}
   \sfcap{Left panel: analysed surface current field (assimilation). Right
   panel: Surface current field of the freerunning model (no assimilation)}
   \label{fig:ass}
\end{figure}

\begin{figure}[h]
   \begin{center}
   \centerline{
   \begin{tabular}{ll}
      \begin{sideways}
      \epsfig{file=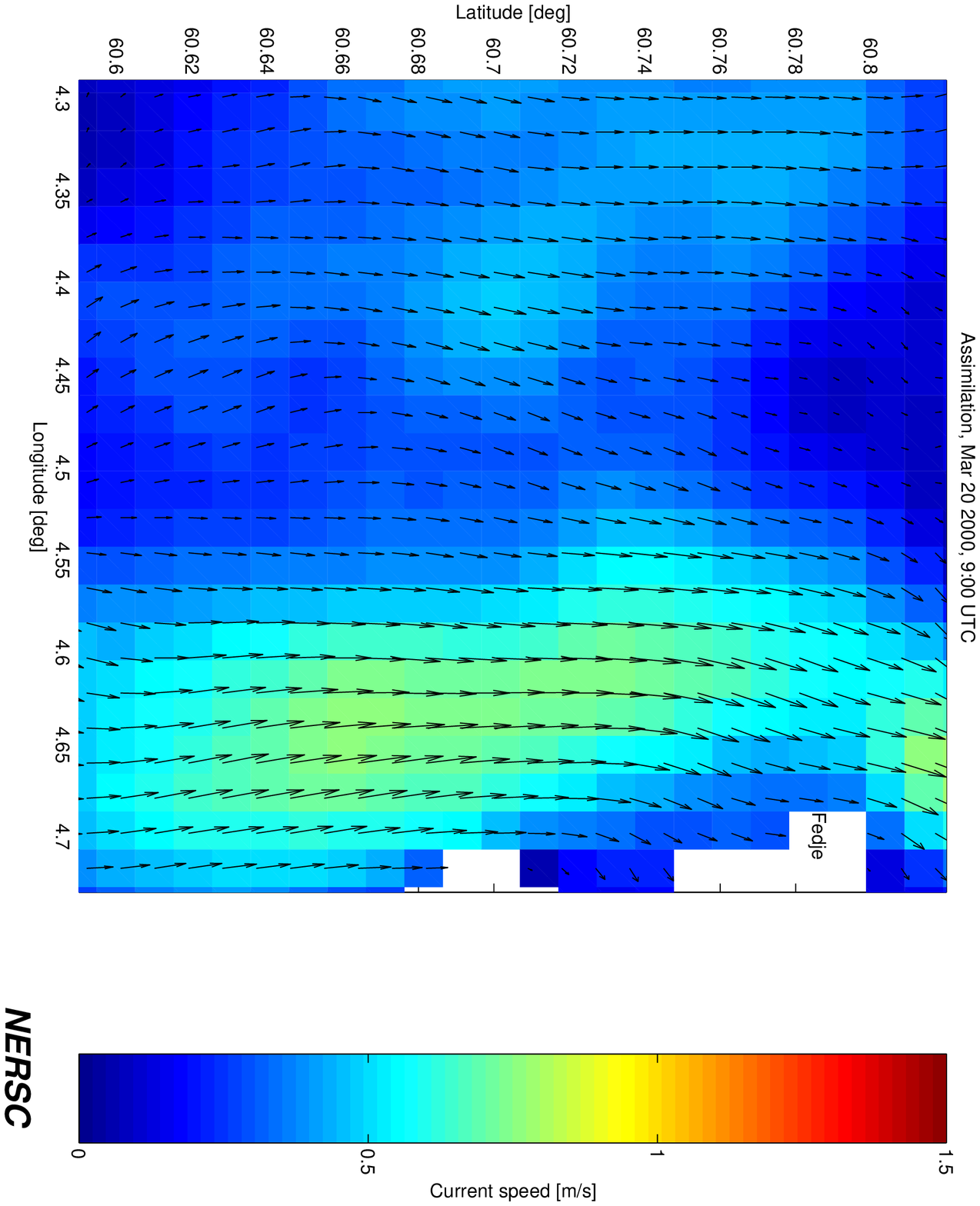, width=0.5\textwidth} \hspace{-0.7cm}
      \end{sideways}
    & \begin{sideways}
      \epsfig{file=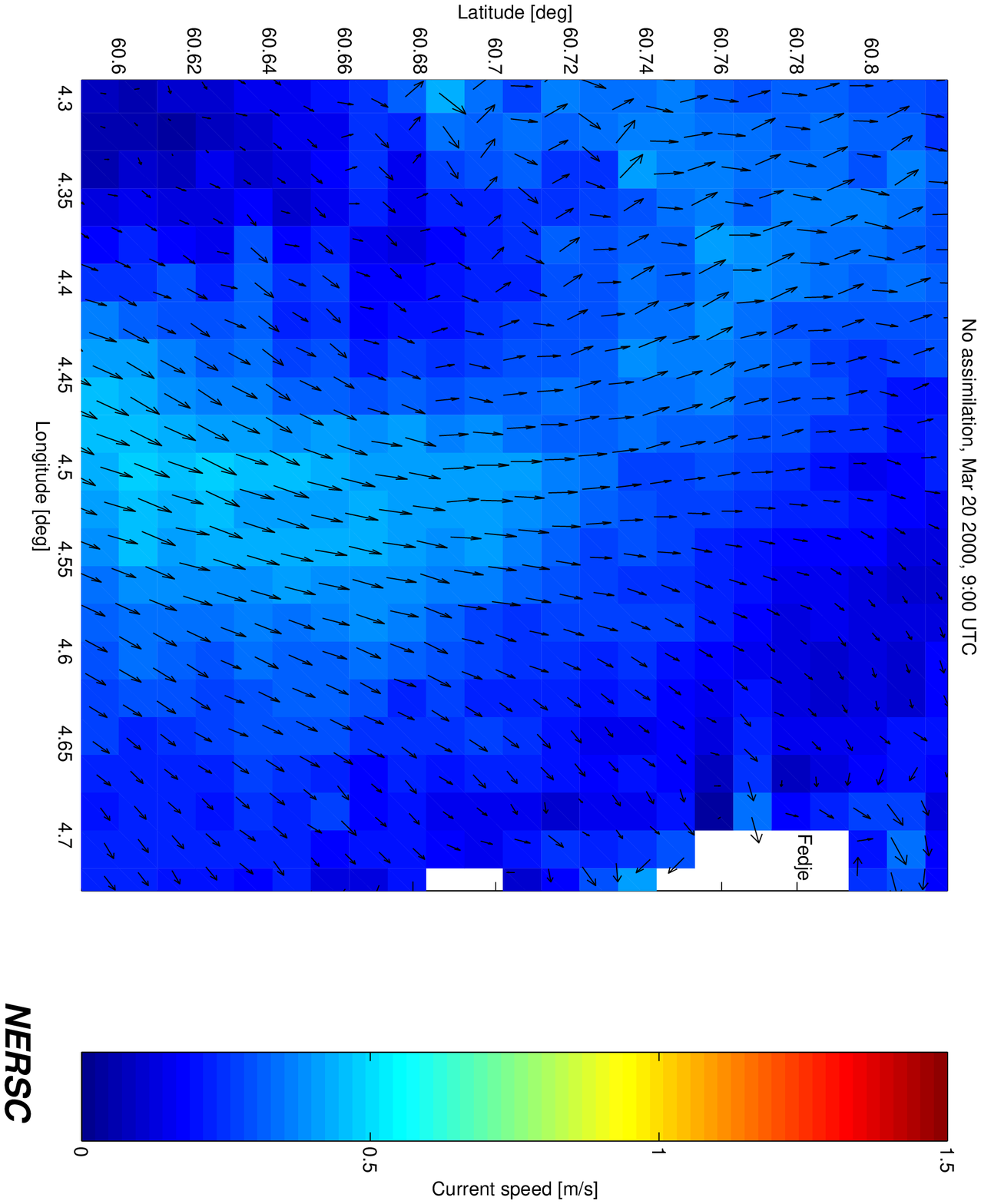, width=0.5\textwidth}
      \end{sideways}
   \end{tabular} }
   \end{center}
   \begin{center}
   \centerline{
      \vspace{-1.0cm} 
      \begin{sideways}
      \epsfig{file=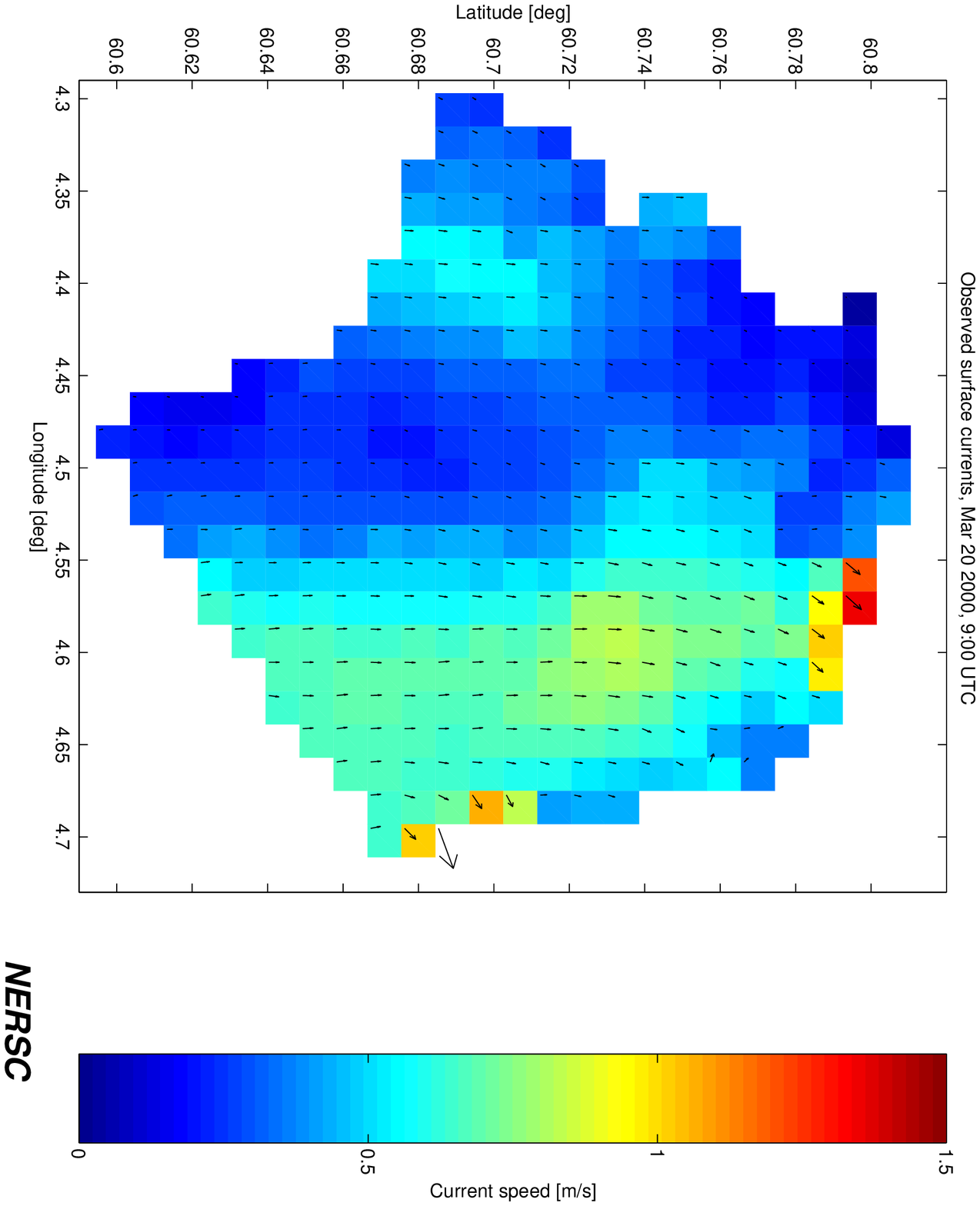, width=0.5\textwidth} \vspace{-1.5cm}
      \end{sideways} }
   \end{center}
   \sfcap{Same as previous figure zoomed in on the area covered by the 
    radar. Radar data are shown in the lower panel for comparison. Arrows are
    only indicative of direction, colour scale gives speed.}
   \label{fig:zoom}
\end{figure}

For an example of the difference between the free run and the analysis, compare
the left and right panels of \Fig{ass}. The free run is clearly less energetic,
and the coastal current appears wider and more diffuse than the assimilated
current field. \Fig{zoom} shows a close-up of the
radar covered area.  It is obvious that the assimilation scheme captures the
strength and extent of the coastal current very well in this particular instant.

More important than the quality of the analysis itself is the temporal impact
of the analysis. I.e., for how long does the added information keep the model
forecast on track?  To assess this, we computed the spatially averaged kinetic
energy in the radar covered area for both model and observations;
\begin{equation}
  \Mean{E}_{\text{k}} = \frac{1}{\Omega} \int_\Omega U^2 \, ds
\end{equation}
where $\Omega$ is the area covered by the radar and $U \equiv
||\vek{u}_\mathrm{H}||$ is the horizontal current speed. The density is assumed
constant and left out. This method has the advantage of appreciating a
corrected current field (typically the coastal current) even when the current
maximum is slightly dislocated by the model yet still improved over the free
run.

\begin{figure}[h]
 \centerline{
  \epsfig{file=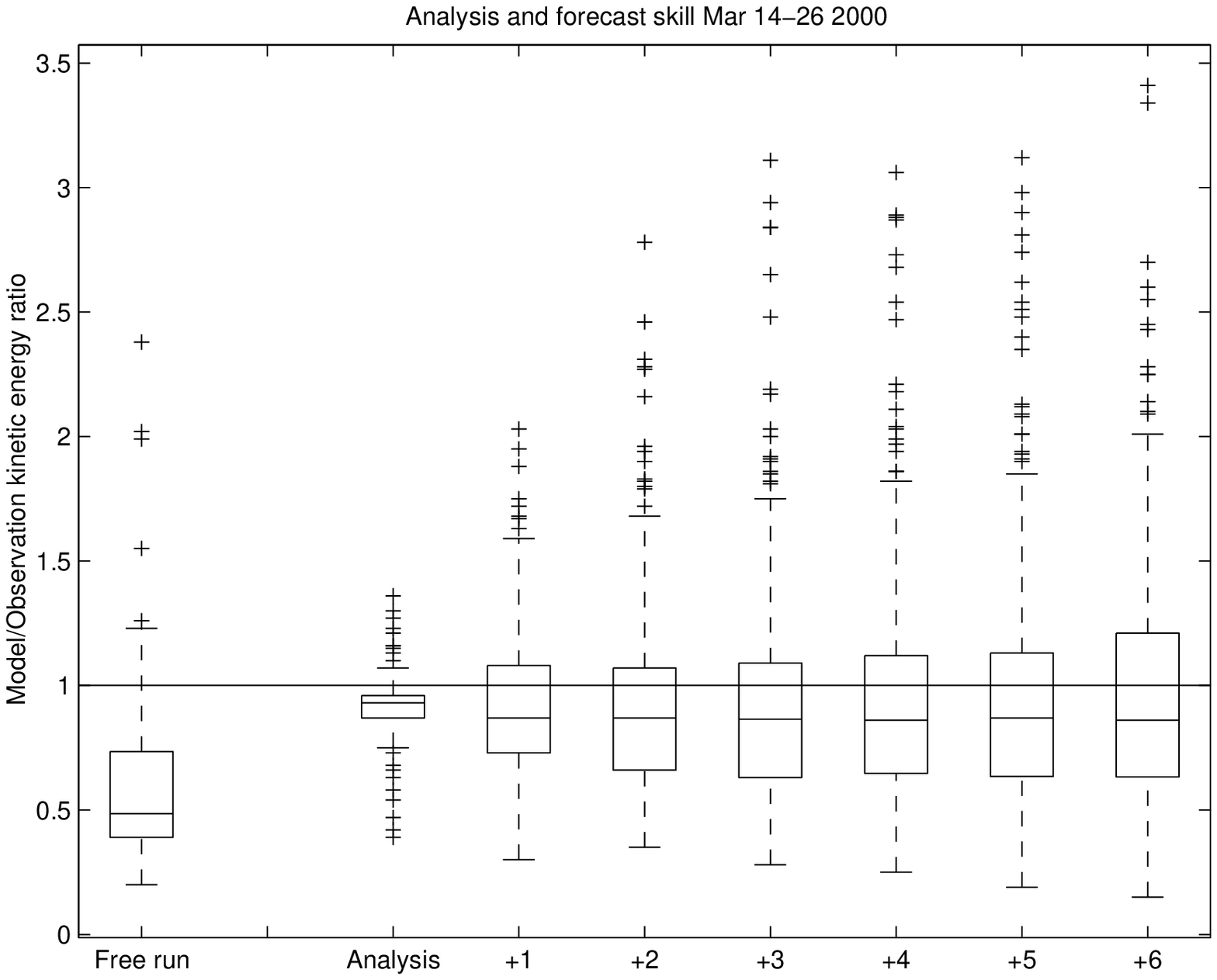, width=0.8\textwidth}
 }
 \sfcap{The assimilation skill measured in level of kinetic energy.
 The ratio of modelled kinetic energy to observed kinetic energy is plotted
 for the free run model (leftmost boxplot), the analysis, and the forecasts,
 numbered in forecast lead time (1 to 6). The boxplots consist of a
 box covering the middle 50\% of the data (quartile to quartile), the median
 line dividing the box, and whiskers indicating the extent of the remaining
 data. Outliers are plotted as individual crosses.}
 \label{fig:boxplot}
\end{figure}

\Fig{boxplot} shows the ratio of the model fields to the observed fields,
$\Mean{E}_{\text{k}}^{\text{mod}}/\Mean{E}_{\text{k}}^{\text{obs}}$. As can be
seen, the free run underestimates the energy level of the coastal current
(roughly by 50\%). The analysis is a significant improvement over the free run
(same figure), with an energy level on a par with what is observed. After
analysis, the forecasts spread out quickly, but retain an average energy level
well above that for the freerunning model even after a six hour forecast.

\begin{figure}[h]
 \centerline{
  \epsfig{file=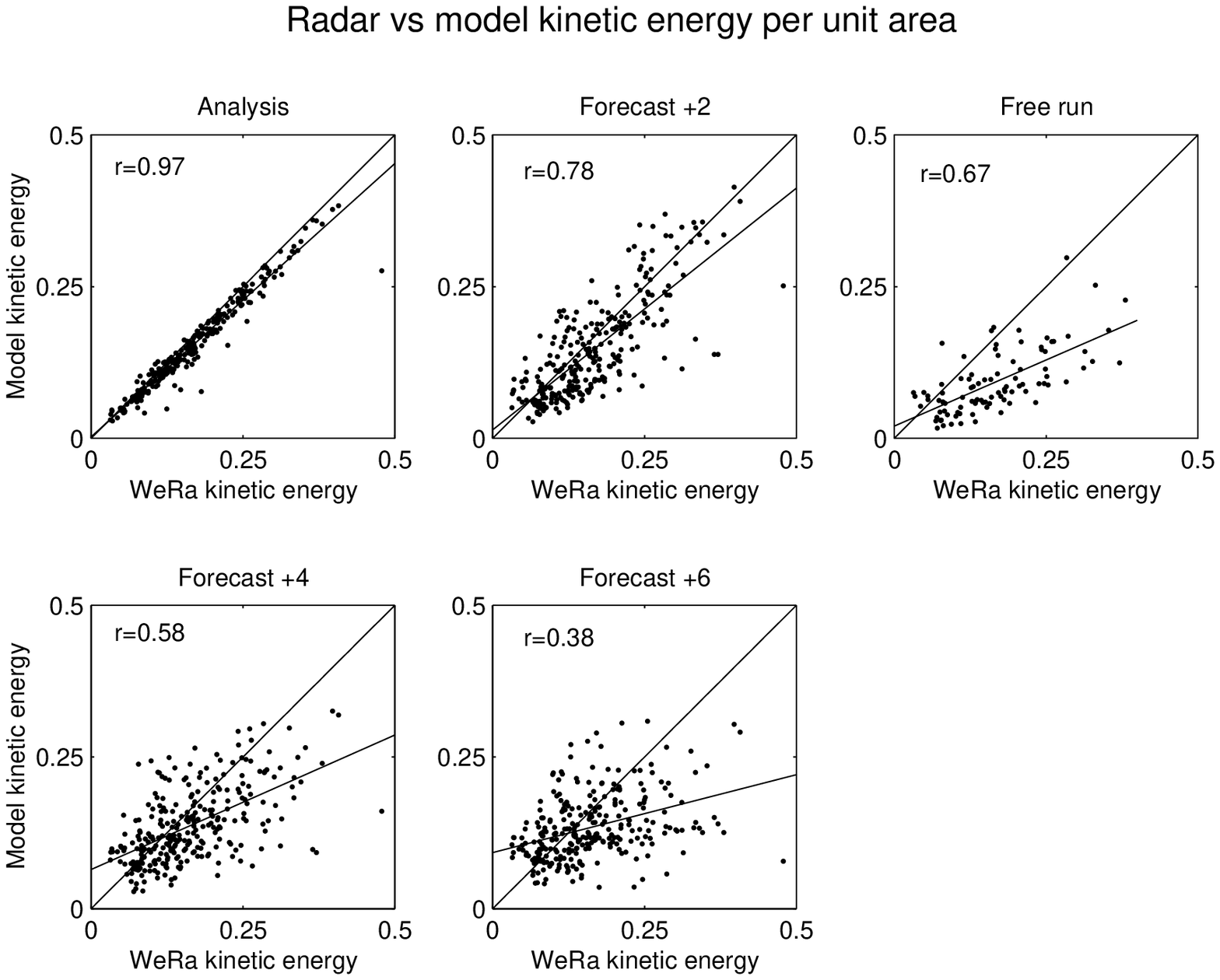, width=0.8\textwidth}
 }
 \sfcap{Scatter plots of observed (WERA) and modelled average energy from the
 assimilation (analysis and forecasts +2 to +6) and the free run. The
 correlation coefficient $r$ is given in the upper left hand corner of each
 panel. The linear regression best fit line is indicated together with the
 ideal $45^\circ$ line.} 
 \label{fig:subenergy} 
\end{figure}

The scatter plot is a more conventional way to compare datasets.
\Fig{subenergy} shows how the energy level decorrelates as the forecast time
increases. At the time of analysis (the assimilation itself), the fields match
up almost perfectly. Two hours from analysis, the correlation is still good,
but as the boxplots also suggested, there is a lot more spread. Another
observation is that the best fit linear regression line dips down compared to
the ideal $45^\circ$ line, which means that the energy level drops off with
forecast lead time. Compared to the free run, it seems that forecasts up to
three hours correlate better with observations.

The probability distributions of two datasets can be compared with an empirical
quantile-quantile plot (EQQ, see \npcite{kle80} or \npcite{wil95}). The
cumulative distribution function (CDF) of a dataset $X$ is
\begin{equation}
   P_X(t) \equiv \int_{-\infty}^t p_X(t')\,dt'.
\end{equation}
The CDF is a monotonically increasing function, hence its inverse $Q_X \equiv
\inv{P}_X$ may be found. The median (mid point) is $t_{0.5} = Q_X(0.5) =
\inv{P}_X(0.5)$, the upper quartile is $\inv{P}_X(0.75)$, and so forth.
Plotting these values against each other reveal differences in the underlying
probability distributions. If a population $Y$ is a linear transform of 
$X$, i.e.\ $Y = aX +b$, the quantiles will fall on a straight line.

\begin{figure}[h]
 \centerline{
  \epsfig{file=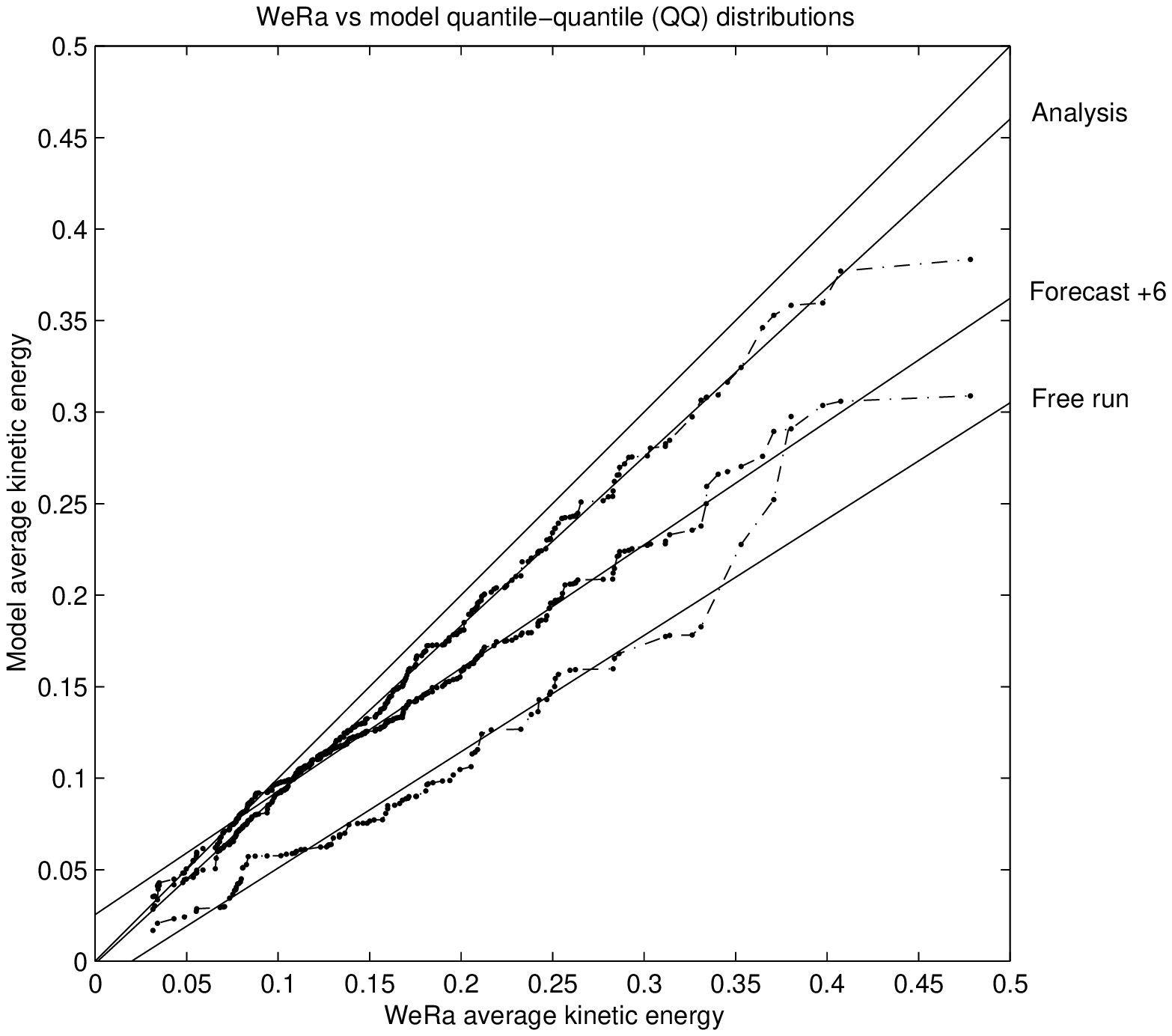, width=0.8\textwidth}
 }
 \sfcap{Empirical quantile-quantile plots of observed (WERA) and modelled
 average kinetic energy from the assimilation (analysis and forecast +6) and
 the free run.  The linear regression best fit lines are indicated together
 with the ideal $45^\circ$ line.}
 \label{fig:qqenergy} 
\end{figure}

\Fig{qqenergy} shows the QQ-plots of average kinetic energy of the analysis,
the +6 hour forecast, and the free run vs radar data. The distribution of the
analysis is almost identical to the radar data, with a slope close to unity.
The +6 hour forecast has a weaker slope, again suggesting that the forecasts
are not able to keep up the coastal current. However, the underlying
distribution seems to be of the same form as the radar data. Finally, the free
run displays an even weaker energy field and stronger deviations from the radar
distribution.  All three datasets display some deviations in the upper extreme
tail.

\begin{figure}[h]
 \centerline{
  \epsfig{file=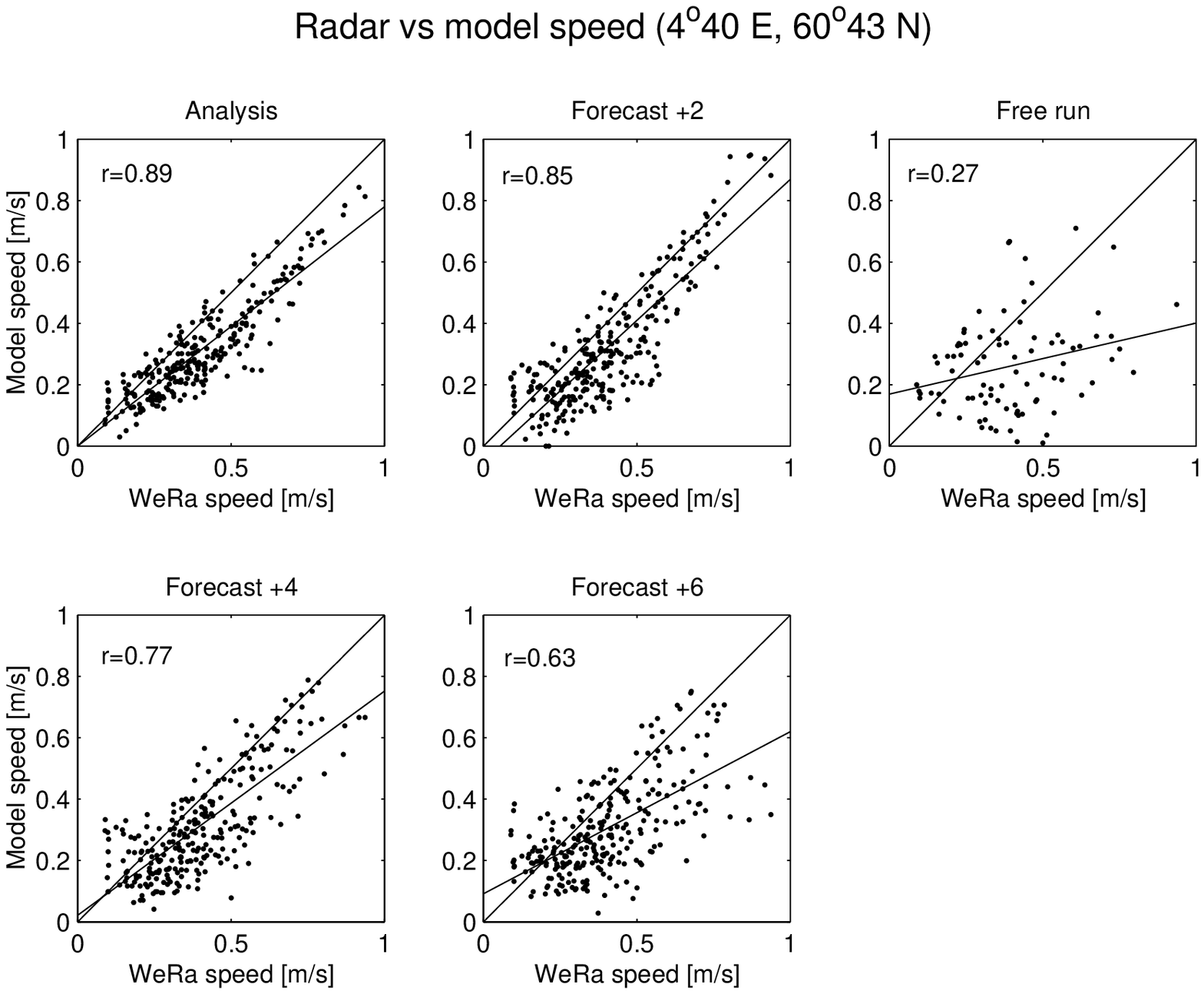, width=0.8\textwidth}
 }
 \sfcap{Scatter plots of observed (WERA) and modelled current speed from the 
 assimilation (analysis and forecasts +2 to +6) and the free run. The 
 correlation coefficient $r$ is given in the upper left hand corner of each 
 panel. The linear regression best fit line is indicated together with the
 ideal $45^\circ$ line.}
 \label{fig:subspeed}
\end{figure}

Time series of current speed were extracted from the radar data and the model
grid point nearest the position $(4^\circ 40'\, \text{E},60^\circ 43'\,
\text{N})$.  This point was chosen because it is in the central area of
interest for the VTS, and always covered by the radar, even in situations with
very low radar coverage. \Fig{subspeed} shows the correlation between observed
and modelled current speed. The freerunning model correlates extremely poorly
with observations in this particular point and is clearly not able to capture
the intensity of the coastal current.

\begin{figure}[h]
 \centerline{
  \epsfig{file=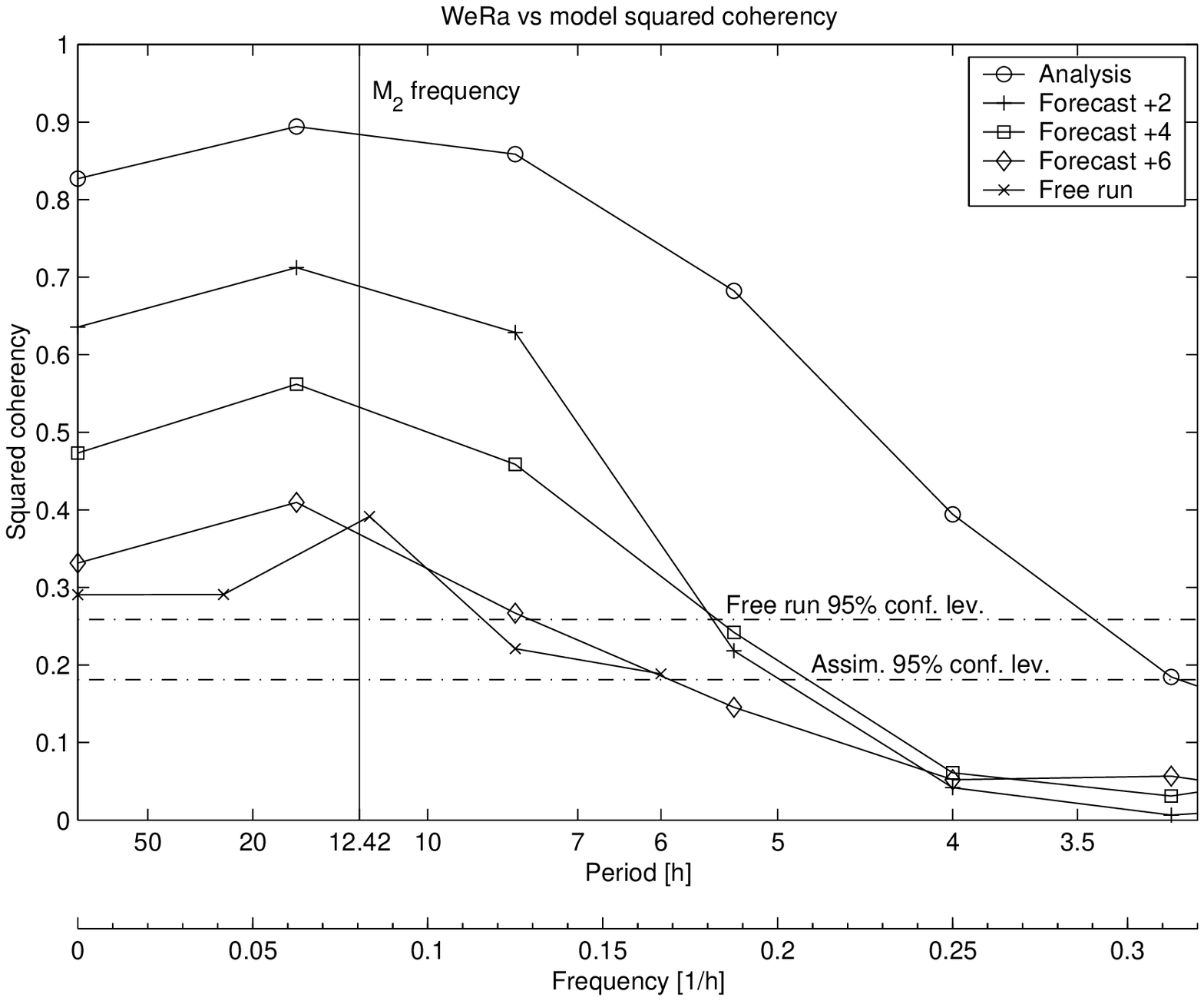, width=0.8\textwidth}
 }
 \sfcap{The squared coherency spectra of observed speed vs model speed time
  series. The dominant tidal constituent $\text{M}_2$ is given for reference
  together with the confidence limits for assimilation data (analysis and
  forecasts +2 to +6) and the free run. Note that the free run has a high
  frequency cutoff due to different sampling rate (every third hour instead of
  every hour as for the assimilation). Frequencies are given in
  $[\text{hours}]^{-1}$.}

 \label{fig:cohere}
\end{figure}

Finally, the spectral characteristics of the modelled currents were studied.
Time series of analysed speed, forecasted speed (at different lead times), and
forecasts from the free run were compared to the radar data.  The time series
were subjected to a coherency analysis. The squared coherency spectrum between
two time series is defined as
\begin{equation}
   \gamma^2(f) \equiv \frac{|G_{12}(f)|^2}{G_{11}(f)G_{22}(f)},
   \label{eq:cohere}
\end{equation}
where $G_{11}$ and $G_{22}$ denote the power spectral density (variance
spectra) of the observed and modelled time series, and $G_{12}$ is the complex
cross spectral density (see \npcite{eme97}).  Confidence limits based on the
number of equivalent degrees of freedom $n$ are computed following
\incite{tho79},
\begin{equation}
   c^2 = 1-\alpha^{1/(n-1)}.
   \label{eq:conf}
\end{equation}
Here, $\alpha$ indicates confidence level ($\alpha = 0.05$ means a 95\%
confidence interval). The limiting value $c^2$ gives the level up to which
squared coherency values may occur by chance \cite{eme97}. \Fig{cohere} shows
the squared coherency for the analysis, the free run, and forecast lead times
+2, +4, and +6 hours. Good coherency is found for the analysis. The coherency
then drops regularly with forecast lead time until it becomes barely
distinguishable from the free run at six hours from the analysis time.  The
forecasts cross the confidence limit at approximately five-hourly periods ($f
\approx 0.2 \,\text{h}^{-1}$), hence higher frequencies are not trustworthy.
For the free run, a lower sampling frequency was used, leading to a lower
cutoff frequency compared to analysis and forecasts in the figure. This lower
cutoff also results in a higher (i.e.\, poorer) confidence level for the free
run. For all forecast lead times, we observe a maximum in coherency around the
dominant tidal period $\text{M}_2$, indicating that the tidal motion is also
improved by the assimilation.

We conclude that lower frequencies corresponding to periods of more than 6
hours are improved by assimilating radar data whereas the higher frequencies
are much more difficult to get right. This means that the high frequency
features assimilated into the model do not evolve correctly. This seems
reasonable, as the low frequencies associated with the slow meandering of the
coastal current will be a more persistent feature in the radar data than the
swift, smaller scale eddies that move in and out of the radar view.

\section{Conclusion}
We have demonstrated that it is possible to make real time analyses
and forecasts of coastal currents using a suite of nested ocean models and
continuous radar coverage. Both analysis and forecasts clearly outperform the
free run, indicating that the assimilation has added information to the model,
but the +6 hour forecast is only marginally better (if better at all) than the
free run. The assimilation scheme also improves the spectral characteristics of
the ocean model, especially for frequencies corresponding to periods longer
than five hours.

Given the relatively limited coverage of the HF radar, the analyses provide
valuable added information through the extrapolation from radar observations to
the surrounding waters covered by the model grid. 

We were unable to correct the hydrography using the radar currents. This
appears to be a fundamental problem with assimilation in nested models. The
only way to avoid this will be to do assimilation in both the outer and the
inner model. This is a much more time-consuming approach and was not considered
feasible in this real time experiment. We also conclude that the relatively weak
cross-correlations observed between modelled surface currents and the
hydrography discourage this line of approach.

In general, the assimilation scheme is sufficiently sophisticated to allow for
long-ranging corrections outside the actual radar coverage (extrapolation), yet
fast enough to fit in the tight schedule of a real time framework. The total
time from acquisition of data until the presentation of analysis and
forecast was ready at the Vessel Traffic Service in Fedje was 45 min. The
system was found to be quite robust to bad observations and was able to operate
during periods of high and low radar coverage.

\begin{flushleft}
\emph{Acknowledgements}\\
The authors were funded by the EU MAST-III project ``European Radar Ocean
Sensing'' (EuroROSE), contract no. PL971607. The authors wish to thank
Klaus-Werner Gurgel for delivering the radar data and Geir Evensen for
discussions on assimilation and diagnosing the results. Thanks also to
Heinz G\"{u}nther for coordinating EuroROSE. The thoughtful reviews by
Dr Jun She and one anonymous reviewer are also appreciated.
\end{flushleft}


{\clearpage}
\bibliographystyle{apacite}
\bibliography{/home/rd/diob/Doc/TeX/Bibtex/BreivikAbb,/home/rd/diob/Doc/TeX/Bibtex/Breivik}

\begin{thebibliography}{}

\bibitem[\protect\citeauthoryear{%
Barrick%
}{%
Barrick%
}{%
{\protect\APACyear{1978}}%
}]{%
bar78}%
\APACinsertmetastar{%
bar78}%
Barrick, D\BPBI E.%
%
\unskip\
\newblock
\APACrefYearMonthDay{1978}{}{}.
\newblock
\BBOQ{}\APACrefatitle{{HF Radio Oceanography---A Review}}{{HF Radio
  Oceanography---A Review}}.\BBCQ{}
\newblock
\APACjournalVolNumPages{Boundary Layer Meteorol}{13}{}{23--43}.
\PrintBackRefs{\CurrentBib}

\bibitem[\protect\citeauthoryear{%
Barrick%
, Evans%
\BCBL{}\ \BBA{} Weber%
}{%
Barrick%
\ \protect\BOthers{.}}{%
{\protect\APACyear{1977}}%
}]{%
bar77}%
\APACinsertmetastar{%
bar77}%
Barrick, D\BPBI E.%
, Evans, M\BPBI W.%
\BCBL{}\ \BBA{} Weber, B\BPBI L.%
%
\unskip\
\newblock
\APACrefYearMonthDay{1977}{}{}.
\newblock
\BBOQ{}\APACrefatitle{{Ocean Surface Currents Mapped by Radar}}{{Ocean Surface
  Currents Mapped by Radar}}.\BBCQ{}
\newblock
\APACjournalVolNumPages{Science}{198}{}{138--144}.
\PrintBackRefs{\CurrentBib}

\bibitem[\protect\citeauthoryear{%
Blumberg%
\ \BBA{} Mellor%
}{%
Blumberg%
\ \BBA{} Mellor%
}{%
{\protect\APACyear{1987}}%
}]{%
blu87}%
\APACinsertmetastar{%
blu87}%
Blumberg, A\BPBI F.%
\BCBT{}\ \BBA{} Mellor, G\BPBI L.%
%
\unskip\
\newblock
\APACrefYearMonthDay{1987}{}{}.
\newblock
\BBOQ{}\APACrefatitle{{A description of a three-dimensional coastal ocean
  circulation model}}{{A description of a three-dimensional coastal ocean
  circulation model}}.\BBCQ{}
\newblock
\BIn{} N\BPBI S.~Heaps\ (\BED), \APACrefbtitle{Three-Dimensional Coastal Ocean
  Models.}{Three-dimensional coastal ocean models.}
\newblock
\APACaddressPublisher{}{American Geophysical Union, Washington D~C}.
\PrintBackRefs{\CurrentBib}

\bibitem[\protect\citeauthoryear{%
Burgers%
, {van Leeuwen}%
\BCBL{}\ \BBA{} Evensen%
}{%
Burgers%
\ \protect\BOthers{.}}{%
{\protect\APACyear{1998}}%
}]{%
bur98}%
\APACinsertmetastar{%
bur98}%
Burgers, G.%
, {van Leeuwen}, P\BPBI J.%
\BCBL{}\ \BBA{} Evensen, G.%
%
\unskip\
\newblock
\APACrefYearMonthDay{1998}{}{}.
\newblock
\BBOQ{}\APACrefatitle{{Analysis Scheme in the Ensemble Kalman
  Filter}}{{Analysis Scheme in the Ensemble Kalman Filter}}.\BBCQ{}
\newblock
\APACjournalVolNumPages{Mon Wea Rev}{126}{}{1719--1724}.
\PrintBackRefs{\CurrentBib}

\bibitem[\protect\citeauthoryear{%
Chapman%
\ \protect\BOthers{.}}{%
Chapman%
\ \protect\BOthers{.}}{%
{\protect\APACyear{1997}}%
}]{%
cha97}%
\APACinsertmetastar{%
cha97}%
Chapman, R\BPBI D.%
, Shay, L\BPBI K.%
, Graber, H\BPBI C.%
, Edson, J\BPBI B.%
, Karachintsev, A.%
, Trump, C\BPBI L.%
\BCBL{}\ \BOthersPeriod{.}%
\unskip\
\newblock
\APACrefYearMonthDay{1997}{}{}.
\newblock
\BBOQ{}\APACrefatitle{{On the accuracy of HF radar surface current
  measurements: Intercomparisons with ship-based sensors}}{{On the accuracy of
  HF radar surface current measurements: Intercomparisons with ship-based
  sensors}}.\BBCQ{}
\newblock
\APACjournalVolNumPages{J Geophys Res}{102}{C8}{18737--18748}.
\PrintBackRefs{\CurrentBib}

\bibitem[\protect\citeauthoryear{%
Daley%
}{%
Daley%
}{%
{\protect\APACyear{1991}}%
}]{%
dal91}%
\APACinsertmetastar{%
dal91}%
Daley, R.%
%
\unskip\
\newblock
\APACrefYear{1991}.
\newblock
\APACrefbtitle{{Atmospheric Data Analysis}}{{Atmospheric Data Analysis}}\
  (\BVOL~2).
\newblock
\APACaddressPublisher{Cambridge}{Cambridge University Press}.
\PrintBackRefs{\CurrentBib}

\bibitem[\protect\citeauthoryear{%
Emery%
\ \BBA{} Thomson%
}{%
Emery%
\ \BBA{} Thomson%
}{%
{\protect\APACyear{1997}}%
}]{%
eme97}%
\APACinsertmetastar{%
eme97}%
Emery, W\BPBI J.%
\BCBT{}\ \BBA{} Thomson, R\BPBI E.%
%
\unskip\
\newblock
\APACrefYear{1997}.
\newblock
\APACrefbtitle{{Data analysis methods in physical oceanography}}{{Data analysis
  methods in physical oceanography}}.
\newblock
\APACaddressPublisher{New York}{Pergamon Press}.
\PrintBackRefs{\CurrentBib}

\bibitem[\protect\citeauthoryear{%
Engedahl%
}{%
Engedahl%
}{%
{\protect\APACyear{1995}}%
}]{%
eng95}%
\APACinsertmetastar{%
eng95}%
Engedahl, H.%
%
\unskip\
\newblock
\APACrefYearMonthDay{1995}{}{}.
\newblock
\APACrefbtitle{{Implementation of the Princeton Ocean Model (POM/ECOM3D) at The
  Norwegian Meteorological Institute (DNMI)}}{{Implementation of the Princeton
  Ocean Model (POM/ECOM3D) at The Norwegian Meteorological Institute (DNMI)}}\
  \APACbVolEdTR{}{Research Report\ \BNUM~5}.
\newblock
\APACaddressInstitution{Oslo, Norway}{The Norwegian Meteorological Institute}.
\PrintBackRefs{\CurrentBib}

\bibitem[\protect\citeauthoryear{%
Essen%
, Gurgel%
\BCBL{}\ \BBA{} Schlick%
}{%
Essen%
\ \protect\BOthers{.}}{%
{\protect\APACyear{2000}}%
}]{%
ess00}%
\APACinsertmetastar{%
ess00}%
Essen, H\BHBI H.%
, Gurgel, K\BHBI W.%
\BCBL{}\ \BBA{} Schlick, T.%
%
\unskip\
\newblock
\APACrefYearMonthDay{2000}{}{}.
\newblock
\BBOQ{}\APACrefatitle{{On the accuracy of current measurements by means of HF
  radar}}{{On the accuracy of current measurements by means of HF
  radar}}.\BBCQ{}
\newblock
\APACjournalVolNumPages{IEEE J~Ocean Eng}{25}{4}{472--480}.
\PrintBackRefs{\CurrentBib}

\bibitem[\protect\citeauthoryear{%
Evensen%
}{%
Evensen%
}{%
{\protect\APACyear{1994}}%
}]{%
eve94a}%
\APACinsertmetastar{%
eve94a}%
Evensen, G.%
%
\unskip\
\newblock
\APACrefYearMonthDay{1994}{}{}.
\newblock
\BBOQ{}\APACrefatitle{{Sequential data assimilation with a nonlinear
  quasi-geostrophic model using Monte Carlo methods to forecast error
  statistics}}{{Sequential data assimilation with a nonlinear quasi-geostrophic
  model using Monte Carlo methods to forecast error statistics}}.\BBCQ{}
\newblock
\APACjournalVolNumPages{J Geophys Res}{99}{C5}{10143--10162}.
\PrintBackRefs{\CurrentBib}

\bibitem[\protect\citeauthoryear{%
Evensen%
}{%
Evensen%
}{%
{\protect\APACyear{1997}}%
}]{%
eve97a}%
\APACinsertmetastar{%
eve97a}%
Evensen, G.%
%
\unskip\
\newblock
\APACrefYearMonthDay{1997}{}{}.
\newblock
\BBOQ{}\APACrefatitle{{Advanced data assimilation for strongly nonlinear
  dynamics}}{{Advanced data assimilation for strongly nonlinear
  dynamics}}.\BBCQ{}
\newblock
\APACjournalVolNumPages{Mon Wea Rev}{125}{}{1342-1354}.
\PrintBackRefs{\CurrentBib}

\bibitem[\protect\citeauthoryear{%
Fernandez%
, Vesecky%
\BCBL{}\ \BBA{} Teague%
}{%
Fernandez%
\ \protect\BOthers{.}}{%
{\protect\APACyear{1996}}%
}]{%
fer96}%
\APACinsertmetastar{%
fer96}%
Fernandez, D\BPBI M.%
, Vesecky, J\BPBI F.%
\BCBL{}\ \BBA{} Teague, C\BPBI C.%
%
\unskip\
\newblock
\APACrefYearMonthDay{1996}{}{}.
\newblock
\BBOQ{}\APACrefatitle{{Measurements of upper ocean surface current shear with
  high-frequency radar}}{{Measurements of upper ocean surface current shear
  with high-frequency radar}}.\BBCQ{}
\newblock
\APACjournalVolNumPages{J Geophys Res}{101}{C12}{28615--28625}.
\PrintBackRefs{\CurrentBib}

\bibitem[\protect\citeauthoryear{%
Gill%
}{%
Gill%
}{%
{\protect\APACyear{1982}}%
}]{%
gil82}%
\APACinsertmetastar{%
gil82}%
Gill, A\BPBI E.%
%
\unskip\
\newblock
\APACrefYear{1982}.
\newblock
\APACrefbtitle{{Atmosphere-Ocean Dynamics}}{{Atmosphere-Ocean Dynamics}}\
  (\BVOL~30).
\newblock
\APACaddressPublisher{San Diego, California}{Academic Press, Inc}.
\PrintBackRefs{\CurrentBib}

\bibitem[\protect\citeauthoryear{%
Glenn%
, Boicourt%
, Parker%
\BCBL{}\ \BBA{} Dickey%
}{%
Glenn%
\ \protect\BOthers{.}}{%
{\protect\APACyear{2000}}%
}]{%
gle00}%
\APACinsertmetastar{%
gle00}%
Glenn, S\BPBI M.%
, Boicourt, W.%
, Parker, B.%
\BCBL{}\ \BBA{} Dickey, T\BPBI D.%
%
\unskip\
\newblock
\APACrefYearMonthDay{2000}{}{}.
\newblock
\BBOQ{}\APACrefatitle{{Operational Observation Networks for Ports, a Large
  Estuary and an Open Shelf}}{{Operational Observation Networks for Ports, a
  Large Estuary and an Open Shelf}}.\BBCQ{}
\newblock
\APACjournalVolNumPages{Oceanography}{13}{1}{12--23}.
\PrintBackRefs{\CurrentBib}

\bibitem[\protect\citeauthoryear{%
Gurgel%
\ \BBA{} Antonischki%
}{%
Gurgel%
\ \BBA{} Antonischki%
}{%
{\protect\APACyear{1997}}%
}]{%
gur97}%
\APACinsertmetastar{%
gur97}%
Gurgel, K\BHBI W.%
\BCBT{}\ \BBA{} Antonischki, G.%
%
\unskip\
\newblock
\APACrefYearMonthDay{1997}{}{}.
\newblock
\BBOQ{}\APACrefatitle{{Measurement of surface current fields with high spatial
  resolution by the HF radar WERA}}{{Measurement of surface current fields with
  high spatial resolution by the HF radar WERA}}.\BBCQ{}
\newblock
\BIn{} \APACrefbtitle{Proceedings of the \textsc{IGARSS}'97
  Conference}{Proceedings of the \textsc{IGARSS}'97 conference}\ (\BPGS\
  1820--1822).
\PrintBackRefs{\CurrentBib}

\bibitem[\protect\citeauthoryear{%
Gurgel%
, Antonischki%
, Essen%
\BCBL{}\ \BBA{} Schlick%
}{%
Gurgel%
\ \protect\BOthers{.}}{%
{\protect\APACyear{1999}}%
}]{%
gur99}%
\APACinsertmetastar{%
gur99}%
Gurgel, K\BHBI W.%
, Antonischki, G.%
, Essen, H\BPBI H.%
\BCBL{}\ \BBA{} Schlick, T.%
%
\unskip\
\newblock
\APACrefYearMonthDay{1999}{}{}.
\newblock
\BBOQ{}\APACrefatitle{{Wellen Radar (WERA): a new ground-wave HF radar for
  ocean remote sensing}}{{Wellen Radar (WERA): a new ground-wave HF radar for
  ocean remote sensing}}.\BBCQ{}
\newblock
\APACjournalVolNumPages{Coastal Engineering}{37}{3--4}{219--234}.
\PrintBackRefs{\CurrentBib}

\bibitem[\protect\citeauthoryear{%
Haltiner%
\ \BBA{} Williams%
}{%
Haltiner%
\ \BBA{} Williams%
}{%
{\protect\APACyear{1980}}%
}]{%
hal80}%
\APACinsertmetastar{%
hal80}%
Haltiner, G\BPBI J.%
\BCBT{}\ \BBA{} Williams, R\BPBI T.%
%
\unskip\
\newblock
\APACrefYear{1980}.
\newblock
\APACrefbtitle{{Numerical prediction and dynamic meteorology}}{{Numerical
  prediction and dynamic meteorology}}.
\newblock
\APACaddressPublisher{New York}{Wiley and sons}.
\PrintBackRefs{\CurrentBib}

\bibitem[\protect\citeauthoryear{%
Ikeda%
, Johannessen%
, Lygre%
\BCBL{}\ \BBA{} Sandven%
}{%
Ikeda%
\ \protect\BOthers{.}}{%
{\protect\APACyear{1989}}%
}]{%
ike89}%
\APACinsertmetastar{%
ike89}%
Ikeda, M.%
, Johannessen, J\BPBI A.%
, Lygre, K.%
\BCBL{}\ \BBA{} Sandven, S.%
%
\unskip\
\newblock
\APACrefYearMonthDay{1989}{}{}.
\newblock
\BBOQ{}\APACrefatitle{{A process study of mesoscale meanders and eddies in the
  Norwegian Coastal Current}}{{A process study of mesoscale meanders and eddies
  in the Norwegian Coastal Current}}.\BBCQ{}
\newblock
\APACjournalVolNumPages{J Phys Oceanogr}{19}{}{20--35}.
\PrintBackRefs{\CurrentBib}

\bibitem[\protect\citeauthoryear{%
Johannessen%
, Svendsen%
, Sandven%
, Johannessen%
\BCBL{}\ \BBA{} Lygre%
}{%
Johannessen%
\ \protect\BOthers{.}}{%
{\protect\APACyear{1989}}%
}]{%
joh89}%
\APACinsertmetastar{%
joh89}%
Johannessen, J\BPBI A.%
, Svendsen, E.%
, Sandven, S.%
, Johannessen, O\BPBI M.%
\BCBL{}\ \BBA{} Lygre, K.%
%
\unskip\
\newblock
\APACrefYearMonthDay{1989}{}{}.
\newblock
\BBOQ{}\APACrefatitle{{Three-Dimensional Structure of Mesoscale Eddies in the
  Norwegian Coastal Current}}{{Three-Dimensional Structure of Mesoscale Eddies
  in the Norwegian Coastal Current}}.\BBCQ{}
\newblock
\APACjournalVolNumPages{J Phys Oceanogr}{19}{}{3--19}.
\PrintBackRefs{\CurrentBib}

\bibitem[\protect\citeauthoryear{%
Kleiner%
\ \BBA{} Graedel%
}{%
Kleiner%
\ \BBA{} Graedel%
}{%
{\protect\APACyear{1980}}%
}]{%
kle80}%
\APACinsertmetastar{%
kle80}%
Kleiner, B.%
\BCBT{}\ \BBA{} Graedel, T\BPBI E.%
%
\unskip\
\newblock
\APACrefYearMonthDay{1980}{}{}.
\newblock
\BBOQ{}\APACrefatitle{{Exploratory Data Analysis in the Geophysical
  Sciences}}{{Exploratory Data Analysis in the Geophysical Sciences}}.\BBCQ{}
\newblock
\APACjournalVolNumPages{Rev Geophys Space Phys}{18}{}{699--717}.
\PrintBackRefs{\CurrentBib}

\bibitem[\protect\citeauthoryear{%
Kowalik%
\ \BBA{} Murty%
}{%
Kowalik%
\ \BBA{} Murty%
}{%
{\protect\APACyear{1993}}%
}]{%
kow93}%
\APACinsertmetastar{%
kow93}%
Kowalik, Z.%
\BCBT{}\ \BBA{} Murty, T\BPBI S.%
%
\unskip\
\newblock
\APACrefYear{1993}.
\newblock
\APACrefbtitle{{Numerical modeling of ocean dynamics}}{{Numerical modeling of
  ocean dynamics}}\ (\BVOL~4).
\newblock
\APACaddressPublisher{New York}{World Scientific Publishing}.
\PrintBackRefs{\CurrentBib}

\bibitem[\protect\citeauthoryear{%
Martinsen%
\ \BBA{} Engedahl%
}{%
Martinsen%
\ \BBA{} Engedahl%
}{%
{\protect\APACyear{1987}}%
}]{%
mar87}%
\APACinsertmetastar{%
mar87}%
Martinsen, E\BPBI A.%
\BCBT{}\ \BBA{} Engedahl, H.%
%
\unskip\
\newblock
\APACrefYearMonthDay{1987}{}{}.
\newblock
\BBOQ{}\APACrefatitle{{Implementation and Testing of a Lateral Boundary Scheme
  as an Open Boundary Condition in a Barotropic Ocean Model}}{{Implementation
  and Testing of a Lateral Boundary Scheme as an Open Boundary Condition in a
  Barotropic Ocean Model}}.\BBCQ{}
\newblock
\APACjournalVolNumPages{Coast Eng}{11}{}{603--627}.
\PrintBackRefs{\CurrentBib}

\bibitem[\protect\citeauthoryear{%
Mellor%
\ \BBA{} Yamada%
}{%
Mellor%
\ \BBA{} Yamada%
}{%
{\protect\APACyear{1982}}%
}]{%
mel82}%
\APACinsertmetastar{%
mel82}%
Mellor, G\BPBI L.%
\BCBT{}\ \BBA{} Yamada, T.%
%
\unskip\
\newblock
\APACrefYearMonthDay{1982}{}{}.
\newblock
\BBOQ{}\APACrefatitle{{Development of a turbulent closure model for geophysical
  fluid problems}}{{Development of a turbulent closure model for geophysical
  fluid problems}}.\BBCQ{}
\newblock
\APACjournalVolNumPages{Rev Geophys Space Phys}{20}{}{851--875}.
\PrintBackRefs{\CurrentBib}

\bibitem[\protect\citeauthoryear{%
Mesinger%
\ \BBA{} Arakawa%
}{%
Mesinger%
\ \BBA{} Arakawa%
}{%
{\protect\APACyear{1976}}%
}]{%
mes76}%
\APACinsertmetastar{%
mes76}%
Mesinger, F.%
\BCBT{}\ \BBA{} Arakawa, A.%
%
\unskip\
\newblock
\APACrefYearMonthDay{1976}{}{}.
\newblock
\APACrefbtitle{{Numerical Methods Used in Atmospheric Models}}{{Numerical
  Methods Used in Atmospheric Models}}\ \APACbVolEdTR{}{GARP Publications
  Series\ \BNUM~17}.
\newblock
\APACaddressInstitution{}{World Meteorological Organization}.
\PrintBackRefs{\CurrentBib}

\bibitem[\protect\citeauthoryear{%
Oke%
, Allen%
, Miller%
, Egbert%
\BCBL{}\ \BBA{} Kosro%
}{%
Oke%
\ \protect\BOthers{.}}{%
{\protect\APACyear{2002}}%
}]{%
oke02}%
\APACinsertmetastar{%
oke02}%
Oke, P\BPBI R.%
, Allen, J\BPBI S.%
, Miller, R\BPBI N.%
, Egbert, G\BPBI D.%
\BCBL{}\ \BBA{} Kosro, P\BPBI M.%
%
\unskip\
\newblock
\APACrefYearMonthDay{2002}{}{}.
\newblock
\BBOQ{}\APACrefatitle{{Assimilation of surface velocity data into a primitive
  equation coastal ocean model}}{{Assimilation of surface velocity data into a
  primitive equation coastal ocean model}}.\BBCQ{}
\newblock
\APACjournalVolNumPages{J Geophys Res}{107}{C9}{3122, 25pp,
  doi:10.1029/2000JC000511}.
\PrintBackRefs{\CurrentBib}

\bibitem[\protect\citeauthoryear{%
Richardson%
}{%
Richardson%
}{%
{\protect\APACyear{1922}}%
}]{%
ric22}%
\APACinsertmetastar{%
ric22}%
Richardson, L\BPBI F.%
%
\unskip\
\newblock
\APACrefYear{1922}.
\newblock
\APACrefbtitle{{Weather Prediction by Numerical Process}}{{Weather Prediction
  by Numerical Process}}.
\newblock
\APACaddressPublisher{Cambridge}{Cambridge University Press, reprinted Dover,
  1965}.
\PrintBackRefs{\CurrentBib}

\bibitem[\protect\citeauthoryear{%
Shapiro%
}{%
Shapiro%
}{%
{\protect\APACyear{1970}}%
}]{%
sha70}%
\APACinsertmetastar{%
sha70}%
Shapiro, R.%
%
\unskip\
\newblock
\APACrefYearMonthDay{1970}{}{}.
\newblock
\BBOQ{}\APACrefatitle{{Smoothing, filtering, and boundary effects}}{{Smoothing,
  filtering, and boundary effects}}.\BBCQ{}
\newblock
\APACjournalVolNumPages{Rev Geophys Space Phys}{8}{}{359--387}.
\PrintBackRefs{\CurrentBib}

\bibitem[\protect\citeauthoryear{%
Song%
\ \BBA{} Haidvogel%
}{%
Song%
\ \BBA{} Haidvogel%
}{%
{\protect\APACyear{1994}}%
}]{%
son94}%
\APACinsertmetastar{%
son94}%
Song, Y.%
\BCBT{}\ \BBA{} Haidvogel, D\BPBI B.%
%
\unskip\
\newblock
\APACrefYearMonthDay{1994}{}{}.
\newblock
\BBOQ{}\APACrefatitle{{A semi-implicit ocean circulation model using a
  generalized topography following coordinate system}}{{A semi-implicit ocean
  circulation model using a generalized topography following coordinate
  system}}.\BBCQ{}
\newblock
\APACjournalVolNumPages{J Comput Phys}{115}{}{228--244}.
\PrintBackRefs{\CurrentBib}

\bibitem[\protect\citeauthoryear{%
Thompson%
}{%
Thompson%
}{%
{\protect\APACyear{1979}}%
}]{%
tho79}%
\APACinsertmetastar{%
tho79}%
Thompson, R\BPBI O\BPBI R\BPBI Y.%
%
\unskip\
\newblock
\APACrefYearMonthDay{1979}{}{}.
\newblock
\BBOQ{}\APACrefatitle{{Coherence Significance Levels}}{{Coherence Significance
  Levels}}.\BBCQ{}
\newblock
\APACjournalVolNumPages{J Atmos Sci}{36}{}{2020--2021}.
\PrintBackRefs{\CurrentBib}

\bibitem[\protect\citeauthoryear{%
Wilks%
}{%
Wilks%
}{%
{\protect\APACyear{1995}}%
}]{%
wil95}%
\APACinsertmetastar{%
wil95}%
Wilks, D\BPBI S.%
%
\unskip\
\newblock
\APACrefYear{1995}.
\newblock
\APACrefbtitle{{Statistical Methods in the Atmospheric Sciences}}{{Statistical
  Methods in the Atmospheric Sciences}}.
\newblock
\APACaddressPublisher{London}{Academic Press}.
\PrintBackRefs{\CurrentBib}

\end{thebibliography}

\end{document}